\documentclass{aastex}          
\usepackage{spr-astr-addons}    

\usepackage{color}
\usepackage{comment}
\usepackage{epstopdf}

\usepackage[maxfloats=40] {morefloats} 
\usepackage[misc]{ifsym}

\begin{document}
%
\title{Analysis of selected Kepler Mission planetary light curves}
\shorttitle{Analysis of Kepler light curves}
\shortauthors{<Rhodes \& Budding>}

\author{M.~D.\ Rhodes} 
\affil{BYU, Provo, UT 84602, USA} 
\affil{e-mail: michael\_rhodes@byu.edu}

\author{E.\ Budding }

\affil{ University of Canakkale, TR 17020, Turkey}
\affil{email: ed.budding@xtra.co.nz} 

\affil{E.\ Budding}
\affil{Carter Observatory, Kelburn, Wellington 6012, New Zealand}

\affil{E.\ Budding}
\affil{SCPS, Victoria University of Wellington, PO Box 600, Wellington, New Zealand}

\affil{E.\ Budding}
\affil{Department of Physics and Astronomy, University of Canterbury, ChristChurch 8041, New Zealand}

\begin{abstract}
We have modified the graphical user interfaced close binary system analysis program {\sc CurveFit} to the form {\sc WinKepler} and applied it to 16 representative planetary candidate light curves found in the NASA Exoplanet Archive (NEA) at the Caltech website http://exoplanetarchive.ipac.caltech.edu, with an aim to compare different analytical approaches.  {\sc WinKepler} has parameter options for a realistic physical model, including gravity-brightening and structural parameters derived from the relevant Radau equation. We tested our best-fitting parameter-sets for formal determinacy and adequacy. 

A primary aim is to compare our parameters with those listed in the NEA. Although there are trends of agreement, small differences in the main parameter values are found in some cases, and there may be some relative bias towards a 90\degr\ value for the NEA inclinations. These are assessed against realistic error estimates.

Photometric variability from causes other than planetary transits affects at least 6 of the data-sets studied; with small pulsational behaviour found in 3 of those. For the false positive KOI 4.01, we found that the eclipses could be modelled by a faint background classical Algol as effectively as by a transiting exoplanet. Our empirical checks of limb-darkening, in the cases of KOI 1.01 and 12.01, revealed that the assigned stellar temperatures are probably incorrect. For KOI 13.01, our empirical mass-ratio differs by about 7\% from that of Mislis and Hodgkin (Mon. Not. R. Astron. Soc. 422:1512,2012), who neglected structural effects and higher order terms in the tidal distortion.  Such detailed parameter evaluation, additional to the usual main geometric ones, provides an additional objective for this work.
\end{abstract}


%

\section{Introduction}

The Kepler Mission was launched by NASA in 2009 with the driving aim of discovering Earth-like planets orbiting other stars (Devore et al., 2009). A number of prominent US research institutes are involved in the data recovery and reduction, but the Ames Research Center, in particular, has a central developmental role. The Kepler satellite has continuously surveyed a selected area of about 10$\times$10 degrees in the Cygnus-Lyra region of the Galactic field to determine the proportion of stars, particularly of Main Sequence type, showing planetary transits photometrically. It could be reasonably anticipated that this proportion should not be extremely small (Budding et al., 2005). In fact, from examination of the $\sim$150000 stars in the target list of the Kepler Mission it can be asserted that the fraction of stars showing likely planetary transits is a few per cent. No particular requirement for host star
composition appears required for planetary formation
(Buchave et al., 2012).
 
Kepler's first results were announced in January 2010, with the early planets having relatively short periods. At the time of writing, more than 3000 planetary transits have been recorded and analyzed. Mostly these are attributed to planets larger than the Earth, although about 10\% of candidates hitherto are of a size comparable to that of the Earth. The majority of known examples are smaller than Jupiter, although around 10 percent are of about the same size or larger.  About 5\% have been located in the `habitable zones' of their parent stars.  NASA announced the positive identification of Earth-sized planets towards the end of 2011, and an early survey of the Mission's findings was that of Borucki et al.\ (2011b). It should also be noted  that a fair proportion of initially announced planet finds, perhaps more than $\sim$30\%\ have since been marked as false positives (Matijevic et al., 2012; Bryson et al., 2013).   

The Kepler telescope has a 1.4 m diameter main mirror with fast optics enabling its relatively wide diameter field of view.  There are 21 separate CCD pairs, arranged in a 5$\times$5 square array with the corner elements missing.  Photometric accuracy is related to the position of a stellar image on this array, but for a mean flux count of about 186000 per minute for a mag 12.0 star (cf.\ Table \ref{tab:no2}, below) a measuring standard deviation equivalent to about 383 ppm was estimated by the Kepler Mission engineers (Gilliland et al., 2011). This is close to our finding from the best fitting models pursued in this paper, although certain examples appear to show a wider deviation, for reasons we consider later. 

Software for the primary photometric analysis used to produce the Kepler Mission's findings so far has been developed from, or related to,  modeling procedures for eclipsing binary stars.
A primary aim of the present study is to check the parameters of the selected planetary candidates and host stars found in the NASA Exoplanet Archive (NEA) -- (which for the candidates we have selected are the values found by Batalha et al.\ (2013)  using alternative, or independent, curve-fitting techniques. Given the open access to the NEA, it is useful to compare different methods and check on the closeness of separate modeling results in a way that will add to general confidence on the analysis programs. Regarding exoplanet photometry and its analysis, Burrows (2012) noted the importance of continued awareness of technical developments, while referring also to an established background, citing the well-known Wilson-Devinney (1971) program based on the Roche approximation for binary component shapes. At the same time, Southworth (2012) discussed the merits of the EBOP program (Nelson \& Davis, 1972; Etzel, 1980; Milone, 1993), when developed as a robust and versatile tool for planetary transits.  Classical formulae applying to the eclipses of spherical bodies, but directed specifically to the star and planet configuration, were presented in a 
somewhat generalized form by Mandel and Agol (2002) and applied to the HST light curve of HD209458, while Kang et al.\ (2012) adopted a very simple `box-car' fitting function, but tailored for fast application to large numbers of light curves with the purpose of statistical analysis in mind. Other useful points relevant to  transiting planet photometry were noted by Seager and Mall\'{e}n-Ornelas (2003).

Specifically, in this article, we report on our modifications of the previously produced GUI-based {\sc CurveFit} close binary system analysis program (Zeilik et al., 1988) to the form {\sc WinKepler} and its application to 16 selected planetary candidate light curves from the NEA. A $\chi^2$-minimizing algorithm (Budding \& Najim, 1980; Budding \& Zeilik, 1987; Budding 1993) is built into {\sc WinKepler}.  The `best-fit' model corresponds to the least value of $\chi^2$, defined as $\Sigma(l_{o,i} - l_{c,i})^2/\Delta l_i ^2$ (Bevington, 1969), where $l_{o,i}$ and $l_{c,i}$ are the observed and calculated light levels at a particular phase and $\Delta l_i$ is the error estimate for the measured values of $l_{o,i}$. The NEA provides such estimates in the archive, and we discuss later how this may be checked.

The program contains some user-applied options controlling the approach to the minimum, but, broadly, this follows a Marquardt-Levenberg strategy, with occasional search resettings when an advantageous direction for simultaneous parameter improvements is located.  The formal determinacy of the underlying model can be checked from examination of the error matrix in the vicinity of the optimum fitting.  This error matrix (formed by inversion of the $\chi^2$ Hessian) should be positive definite
for a formally determinate fitting. The eigenvalues and eigenvectors of the $\chi^2$ Hessian are also evaluated as checks on the degree of interdependence of the selected parameter set. Adequacy of the model entails that the value of $\chi^2/\nu$, where $\nu$ is the number of degrees of freedom of the data-set, should be acceptably close to unity at the optimum. If this ratio lies outside, say, a 90\% confidence limit, which can be judged from standard tables of the $\chi^2(\nu)$ distribution, then we can reasonably deduce that there is something amiss. This may be due to some other effect in the data that the model does not take into account, an inappropriately assigned value of $\Delta l$, or failure of the optimization algorithm to converge properly on the best set of parameters.

\begin{figure}[t]
\begin{center}
\includegraphics[width=\columnwidth]{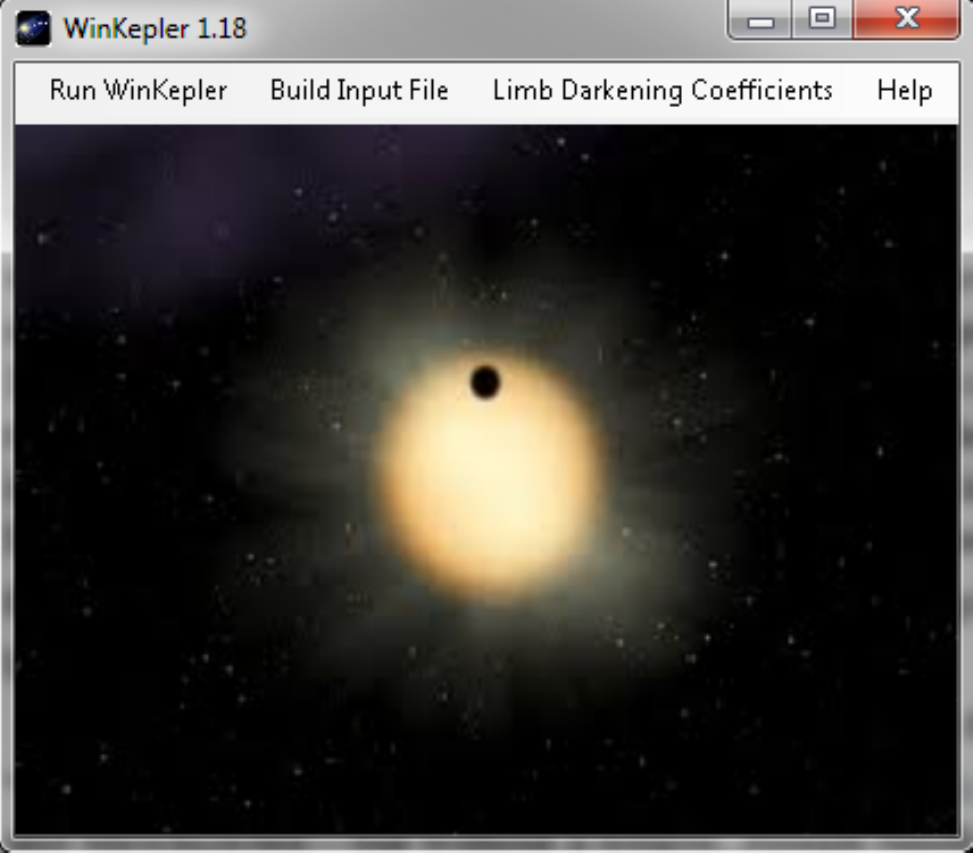} 
\caption{WinKepler Screen Shot}
\label{pic:WinKepler}
\end{center}
\end{figure}

Speed of evaluation of an algebraic form of fitting function is an advantage when exploring a wide range of parameter space and evaluating the parameter error matrix. Relevant experience was gained by the present authors with the Mauna Kea photometry of HD 209458 (cf.\ Budding, Rhodes \& Sullivan, 2005). One interesting question concerned whether there is sufficient information content in the data to permit independent empirical checks on computed models of limb-darkening with sufficient confidence. This can provide an additional objective for the present analysis. {\sc WinKepler} is freeware, and can be downloaded from the website: http://home.comcast.net/{\raise.17ex\hbox{$\scriptstyle\sim$}}michael.rhodes/. A screenshot of {\sc WinKepler}'s  main page is shown in Fig \ref{pic:WinKepler}.

{\sc WinKepler} continues the prescription for tidal and rotational distortions of the components as given by {\sc CurveFit}. This formulation adopts the classical approach to finding the shape of a body distorted by forces associated with rotation and tides by referring to equipotential surfaces. These surfaces can be described in terms of spherical harmonic series. Clauraut's theorem for bodies in equilibrium (cf.\ e.g.\ Pressley, 2001) can then be used to evaluate the coefficients of the terms in the series, with dependence on the structure of the distorted body. Such coefficients ($\eta_j $) have been evaluated for recent models of stars by \.{I}nlek and Budding (2012). For the `first order' solution of Clairaut's equation, in which the interaction of tides on tides is neglected, the structural dependence is summarized by the three tidal terms $w_2= \Delta_2 qr^3 $, $w_3= \Delta_3 qr^4 $, $w_4= \Delta_4 qr^5 $, and the rotational term $v_2= \kappa_{\omega}\Delta_2(1+ q) r^3 $. 
The values of $v$ and $w$ coefficients in this account of the stellar distortion are derived from
numerical integration of a first-order differential equation that Kopal (1959) called Radau's equation. We may therefore refer to the underlying model as the `Radau' model, to distinguish it from others (in the `Roche' model, for example, the foregoing $\Delta_j$s would all revert to unity). The dependent variable in Radau's equation is $\eta_j$: a logarithmic derivative appearing in Clairaut's equation that, in turn, reflects the internal stellar density distribution and is directly related to other parameters (e.g.\ $\Delta_j$, as above, or $\bar{k_j}$) also used in this context. The star's angular rotation $\omega$ is factored in the above terms by  $\kappa_{\omega} = \omega^2 /{\omega_0}^2$; $\omega_0$ corresponding to the default condition ($\kappa_{\omega}= 1$) of synchronized rotation. Further details on {\sc WinKepler} -- its formalism, method and utilization -- are given by Rhodes (2013). Other aspects of the fitting procedure emerge in the applications to individual cases reported on later.

\section{Kepler light curves}

\subsection{Data access and approach to analysis}

The light curve data-sets, consisting mainly of short cadence, normalized PDSCAP\_FLUX values at given times, were downloaded from the NEA.  Details on the light curve selections are found in the later discussion sections for each candidate. Input parameters used to set up the analyses were also downloaded from the NEA, and are shown in Tables \ref{tab:no1} and \ref{tab:no2}). Light curves were selected without any particular preconceptions other than studying a reasonably typical group. In choosing later examples on the list, however, we tended to look for smaller sized planets, in order to find out how well determined their parameters could be. We used only a part of each data-set, that was centered on the minimum and twice the width of the transit duration.  Normally, we did no preliminary binning or detrending. For KOI 42.01, however, where initial values of $\chi^2/\nu$ were relatively high and a clear downward trend could be seen in the residuals across the eclipse, 
we carried out a linear detrending. The issue is discussed in Section 3.12.

Individual transits turn out to have high diagnostic power for the main geometric elements when the data has a reasonably good S/N ratio (the eclipse depth $\sim$10 $\times$ the scatter level, say, and the data-set including at least $\sim$200 points).  But a number of the light curves show additional effects apart from just the planetary transits. Those effects are not pursued in any detail in this paper.   Still, it is necessary to be aware that pulsational microvariability or maculation may be present in the raw data and affect the quality of the transit fittings.  We discuss specific examples in what follows.

%
%

\begin{table*}[]
\small
\caption{Primary Input Data}
\label{tab:no1}
\begin{tabular}{|r|r|r|r|r|r|r|r|r|r|r|}
\tableline 
\multicolumn{1}{|c|}{KOI} & \multicolumn{1}{|c|}{$M_*$} & \multicolumn{1}{|c|}{$R_*$} & \multicolumn{1}{|c|}{$T_{*}$} & \multicolumn{1}{|c|}{$P$} & \multicolumn{1}{|c|}{$Z$} & \multicolumn{1}{|c|}{$\log g$} & \multicolumn{1}{|c|}{$a$} & \multicolumn{1}{|c|}{ ${M_p}/{M_*}$ }& \multicolumn{1}{|c|}{$u$} \\ 

\multicolumn{1}{|c|}{} & \multicolumn{1}{|c|}{($\odot)$} & \multicolumn{1}{|c|}{($\odot$)} & \multicolumn{1}{|c|}{K} & \multicolumn{1}{|c|}{d} & \multicolumn{1}{|c|}{} & \multicolumn{1}{|c|}{(cgs)} & \multicolumn{1}{|c|}{(AU)} & \multicolumn{1}{|c|}{ }& \multicolumn{1}{|c|}{} \\ 
\hline 
\hline
1.01 & 0.995 & 1.06 & 5814 & 2.4706132 & 0.116 & 4.38 & 0.036 & 1E--3 & 0.56 \\ 
\hline 
2.01 & 1.660 & 2.71 & 6264 & 2.2047355 & 0.000 & 3.79 & 0.039 & 5E--3 & 0.51 \\ 
\hline 
3.01 & 0.786 & 0.74 & 4766 & 4.8878003 & 0.000 & 4.59 & 0.052 & 8E-5 & 0.69 \\ 
\hline 
4.01 & 1.610 & 2.60 & 6391 & 3.8493724 & 0.232 & 3.81 & 0.056 & 1E--3 & 0.50 \\ 
\hline 
5.01 & 1.145 & 1.42 & 5861 & 4.7803288 & 0.116 & 4.19 & 0.058 & 1E--6 & 0.55 \\ 
\hline 
7.01 & 1.124 & 1.27 & 5858 & 3.2136641 & --0.085 & 4.28 & 0.044 & 5E--5 & 0.54 \\ 
\hline 
10.01 & 1.140 & 1.56 & 6025 & 3.5224991 & --0.128 & 4.11 & 0.047 & 2E--2 & 0.52 \\ 
\hline 
12.01 & 1.260 & 1.40 & 6419 & 17.8551483 & --0.035 & 4.26 & 0.144 & 1E--3 & 0.50\\ 
\hline 
13.01 & 2.270 & 2.70 & 8848 & 1.7635877 & --0.141 & 3.94 & 0.038 & 3E--3 & 0.44 \\ 
\hline 
17.01 & 1.138 & 1.08 & 5826 & 3.2346996 & 0.000 & 4.42 & 0.045 & 1E--3 & 0.55 \\ 
\hline 
20.01 & 1.172 & 1.38 & 6011 & 4.3796300 & --0.161 & 4.23 & 0.056 & 4E--3 & 0.52 \\ 
\hline 
42.01 & 1.170 & 1.36 & 6170 & 17.834381 & --0.193 & 4.24 & 0.111 & 1E--5 & 0.51 \\ 
\hline 
72.01 & 0.905 & 1.00 & 5627 & 0.8374903 & --0.812 & 4.39 & 0.017 & 1E--6 & 0.50 \\ 
\hline 
117.01 & 1.183 & 1.18 & 5949 & 14.749102 & --0.092& 4.37 & 0.125 & 1E--6 & 0.53 \\ 
\hline 
377.01 & 1.053 & 1.01 & 5777 & 19.2739380 & 0.170 & 4.45 & 0.143 & 2E--6 & 0.57\\ 
\hline 
388.01 & 1.092 & 1.58 & 5707 & 6.1493616 & 0.000 & 4.08 & 0.068 & 4E--7 & 0.56\\ 
\tableline 
\end{tabular} 
\end{table*}

In Table \ref{tab:no1} $M_*$ is the host star's mass (solar masses), $R_*$ -- stellar radius (solar radii), $T_*$ -- effective temperature (K), $P$ –- orbital period (in days), $Z$ -- stellar metallicity (taken to be zero if not given in the source material), $\log g$ -- log$_{10}$(surface gravity) of the star, $a$ -- semi-major axis in AUs, $M_p/M_*$ -- ratio of planet to star masses, $u$ -- stellar (linear) limb-darkening coefficient. All these parameters, with the exception of mass ratio and limb-darkening coefficient, are taken from the NEA.

In making comparisons with parameters published in the NEA website and those from elsewhere it should be noticed that the former were occasionally changed during the period we have been working with the data, and such source information may change again. This may give a temporary quality to any particular set of detailed  numerical results, though this is understandable given the very large ongoing programme of the Kepler Mission. At the present time, we are using the NEA data for the input information relevant to our selected candidates, corresponding to the values listed by Batalha et al (2013).  Even so, comparison of individual curve-fitting results, at any time, reflects the purpose of independent, alternative analysis mentioned before.  The relative scale of the error estimates is also relevant here. The NEA has tended to list parameters always to the same number of decimal places, whereas it is clear that the light curve quality varies considerably from example to example. In many cases, we found that 3 decimal digits suffice to quantify parameter estimates realistically. Sometimes this may reduce to 2 or increase to 4.

The preliminary planetary mass ratios ($M_p/M_*$ --- a formally required fitting input parameter, though usually of negligible effect) in Table \ref{tab:no1} generally come from the archive-listed planet radius, to which a simple scaling from the known mass ratios of solar system planets was applied.\footnote{In the case of KOI 3.01, the mass estimate comes from Bakos et al.\ (2009). This yields a mass $\sim$50\% more than that of the formula.} For planets of greater size, the mean density was assumed equal to that of Jupiter and scaled according to the radius cubed. We derived this quantity and the other parameter at the right of Table \ref{tab:no1}, i.e.\ the coefficient of limb-darkening in the linear approximation $u$, separately, but using the NEA information. For the mass ratio estimate, in view of its essentially approximate nature and very low influence, one significant digit is appropriate, at least initially.

Table \ref{tab:no2} gives further sourced information on the selected examples: the Kepler Input Catalog (KIC) number, an assigned mean planet temperature $T_p$, followed by photometric information, such as the reference out-of-transit flux count used in the fittings. That is a function of the detection system that can be presumed linear with the signal. The depth of the transit is listed in parts per million (ppm). A `Kepler' magnitude has been assigned to accord with the normal astronomical magnitude system,
although the wide bandpass of the detection system complicates its interpretation. We list also a `Poisson factor' (1/$\sqrt{f_{\rm ref}}$). If the mean fluxes of column 4 were actually arrival rates of photons, these numbers would reflect expected errors of measurement.  But the counts are to some extent integrated, scaled and averaged by the detection system's internal electronics. The appropriate error measures for the light curve data would then be these numbers divided by some quantity $s$, say, whose value we derive later.  

The distance $\rho$, in pc (column 7), can be estimated from the formula (Budding \& Demircan, 2007)
\begin{equation}
\log \rho = \log R_* + 0.2V + 2F^{\prime}_V - 7.454  \,\,\,  ,
\end{equation}
where $V$ is the Johnson $V$-magnitude of the star.  $F^{\prime}_V$ is the visual flux parameter defined by
\begin{equation}
F^{\prime}_V  = \log T_e + 0.1BC  \,\,\,  ,
\end{equation}
where $T_e$ is the stellar effective temperature and $BC$ is the bolometric correction, which 
is reasonably well-known for Main Sequence stars of a given $T_e$ (cf.\ Budding \& Demircan, 2007; Table~3.1).
Alternatively, there are accurately known empirical relationships between  $F^{\prime}_V$ and the $V-R$ colour of the star that have been developed since the work of Barnes \& Evans (1976).

 We can consider a transformation from the `Kepler' magnitudes listed in Table \ref{tab:no2} to the standard  $V$ magnitude, since details of the filter used were published by Rowe et al., (2009), but the effects of the wide bandwidth are not small. In this case, a representative conversion formula, using a linearized approximation involving standard magnitudes and colours,  becomes inaccurate for colour indices appreciably different from zero.  On the other hand, the effective wavelength for the relatively small range of near solar temperatures involved does not vary by much from its average value of about 0.61 $\mu$m.   From the given temperatures and corresponding bolometric corrections, we can then use the Barnes-Evans colour correlation in reverse, to derive $V-R$ colours for the candidate stars.  Using the photometric gradient concept (Budding \& Demircan, 2007, Eqn 3.13) and a derived effective wavelength of the Kepler Mission filter for a given effective temperature (Budding \& Demircan, 2007, Eqn 3.30), we can then estimate $V-K$ colours (where $K$ is here the Kepler magnitude) and thence suitable $V$ magnitudes to substitute into Eqn 1. Temperature estimates in the Kepler Input Catalogue have been discussed by Pinsonneault et al (2012), and  some of the original NEA values appear to be in a process of revision. Given also the non-linearity of the relationship of the Kepler magnitudes to monochromatic magnitudes, our derived distances, which neglect interstellar extinction, should be regarded as coarse estimates only. 

\begin{table*}[]
\small
\caption{Additional input information}
\label{tab:no2}
\begin{tabular}{|r|r|r|r|r|r|r|r|}
\tableline 
 \multicolumn{1}{|c|}{KIC} & \multicolumn{1}{|c|}{KOI } & \multicolumn{1}{|c|}{$T_p$}& \multicolumn{1}{|c|}{$f_{\rm ref}$} 
 & \multicolumn{1}{|c|}{Depth } & \multicolumn{1}{|c|}{$K$ mag } & \multicolumn{1}{|c|}{dist.} & \multicolumn{1}{|c|}{Poiss.\ fac.\ }\\ 
 \multicolumn{1}{|c|}{} & \multicolumn{1}{|c|}{} & \multicolumn{1}{|c|}{K}& \multicolumn{1}{|c|}{counts s$^{-1}$} 
 & \multicolumn{1}{|c|}{ppm } & \multicolumn{1}{|c|}{ } & \multicolumn{1}{|c|}{pc} & \multicolumn{1}{|c|}{}\\
\hline 
\hline
11446443 & 1.01 & 1394 & 4.03265E+05 & 14224 & 11.338 & 250 & 0.00157 \\
\hline 
10666592 & 2.01 & 2303 & 1.03525E+06  &6686  & 10.463 &500 & 0.00098 \\
\hline 
10748390 & 3.01 & 794 & 2.88809E+06 &4281 & 9.147 & 41 & 0.00059 \\
\hline 
3861595 & 4.01 &1623  & 3.71398E+05 &1313 & 11.432 &780 & 0.00164  \\
\hline 
8554498 & 5.01 &1279  & 3.19394E+05 & 944 & 11.665 &400 & 0.00177  \\
\hline 
11853905 & 7.01 &1386  & 1.76070E+05 &740  & 12.211 &460 & 0.00238  \\
\hline 
6922244 & 10.01 &1532  & 4.93732E+04 & 9245 & 13.563 &1100 & 0.00450  \\
\hline 
 5812701 & 12.01 &876 &  4.19316E+05 &9271  & 11.353 &400 & 0.00154 \\
\hline 
9941662 & 13.01 &3320  & 1.72255E+06 & 4646 & 9.958 &710 & 0.00076  \\
\hline 
10875245 & 17.01 &1260  & 7.40452E+04&10586  & 13.000 &560 & 0.00367  \\
\hline 
11804465 & 20.01 &1314  & 5.74039E+04&16297  & 13.438 &930 & 0.00417  \\
\hline 
8866102 & 42.01 &845  & 2.68825E+06 & 334 & 9.364 &150 & 0.00061  \\
\hline 
11904151 & 72.01 &1903  & 5.23104E+05 & 191 & 10.961 &190 & 0.00138  \\
\hline 
10875245 & 117.01 & 807 & 1.37001E+05 & 495 & 12.487 &500 & 0.00270 \\
\hline 
 3323887 & 377.01 &678 & 3.61538E+04 &4887  & 13.803 &740 & 0.00509  \\
\hline 
3831053 & 388.01 &1214  & 6.00530E+04 & 362  & 13.644 &1050 & 0.00408  \\
\tableline 
\end{tabular} 
\end{table*}

Of the parameters given in Table \ref{tab:no1}, the orbital period is the most accurately known. At first, we allowed optimization of the period for given data-sets covering several transits, but as more data accumulates this becomes less purposeful, as the technique for determining periodicity is different from fitting phased data. Using a separate procedure to find only orbital ephemerides has an information advantage over deriving these parameters with others from a single fitting function. 

Given the period, if we had also the stellar mass, Kepler's Third Law would yield a fairly reliable separation of planet from star (since we can reasonably neglect the mass of the planet). But prior estimates of the stellar masses came from indirect evidence that depends on additional assumptions. These masses can probably only be estimated to within $\sim$10 percent of their true values. This point has an important bearing on subsequent quantification, since the absolute sizes of objects determined from light curve fittings cannot be more precisely known than the separation of the components.  Although the separation has only a weak dependence (power 1/3) on mass, this uncertainty means that, at best, we could expect only three meaningful digits in the semi-major axes, and usually fewer than three. The values of $a$ listed in Table \ref{tab:no1} and calculated in this way can differ from other possible derivations that we discuss shortly. We have retained the third significant digit in Table \ref{tab:no1} to allow assessment of such differences. 

Differences in the absolute separations of star and planet occur when we use the listed surface gravity and radius values to calculate masses from the formula $\log M = \log g - 4.44 + 2\log R$ (masses and radii in solar units). This formula can produce quite unlikely masses for Main Sequence stars, that are reflected in significantly different $a$ values, despite the low sensitivity to mass. This problem was recognized by Batalha et al (2013), where the observed gravities were `corrected' to those appropriate for Main Sequence stars of the same effective temperature, composition and mean age of the solar neighbourhood. The masses derived from such corrected gravities are the masses given in Table 1.

Inclusion of the gravity constraint, even corrected, will reflect some natural evolutionary spread about the Main Sequence in the listed masses. Here, however, another issue arises, in that the constraint brings in also the stellar radius. This can, in principle, be derived from applying another condition, this time using the photometrically fitted relative radius ($r_1$ in conventional terminology), which allows an estimate of the mean density ($\rho_*$), since $\rho_* = {\rm const.}/(P^2 r_1^3)$. The absolute radius then follows from $R_* = 3g/4\pi G\rho_*$. Note here, though, the sensitivity of $\rho_*$ to the derived $r_1$ value, as well as the imprecisely known gravity. Besides, we can reasonably expect that $\rho_*$ would vary by $\sim$1 order of magnitude for solar type stars on the Main Sequence. The uncertainty thus introduced into the absolute radius must affect masses derived from the gravity constraint pejoratively, and an indication of this was reported by Muirhead et al (2012).

\subsection{Results}

Two different paths were followed to set up our light-curve models: (1) we estimated preliminary parameters from  measurements of light curve segments shown in the NEA display facility, aided by the basic formulae given in Chapter 7 of Budding \& Demircan (2007); (2) we adopted trial parameters from those given in the NEA. Relevant quantities are listed in Table \ref{tab:no3}. 

We have generally optimized the curve-fitting using the five parameters $\Delta \phi_0$, $U$, $r_1$, $k$ and $i$ listed in Table \ref{tab:no3}. The first two of these locate the reference points for the phase and flux axes, respectively. These are of interest for time of mid-transit variation (TTV) studies, or absolute flux calibrations. The other 3 parameters correspond to the usual main geometric `elements'. After each parameter value we give the leading digit of our error estimate as it would affect the last tabulated digit of the parameter value.  Thus, the first entry of Table \ref{tab:no3} means $U = 1.00083 \pm 0.00003$. In will be quickly seen that the light curves vary a great deal in the relative scale of their scatter, with natural consequences for parameter determinability.  In a coarse way, we could thus assign relatively good quality to the light curves of KOI 1.01, 2.01, 3.01, 12.01, 13.01; moderate quality to those of KOI 4.01, 5.01, 10.01, 17.01, 20.01, 377.01; and poor quality to KOI 7.01, 42.01, 72.01, 117.01 and 388.01. Such assessments can guide general expectations on the results.

\begin{table*}[]
\small
\caption{The basic fitting parameters for 16 light curves} 
\label{tab:no3}
\begin{tabular}{|r|l|l|r|l|r|l|r|l|}
\tableline 
\multicolumn{1}{|c|}{KOI} & \multicolumn{1}{|c|}{$U$} & \multicolumn{1}{|c|}{$\Delta \phi_0$} & \multicolumn{1}{|c|}{$r_{1 K}$} & \multicolumn{1}{|c|}{$r_{1 P}$}  & \multicolumn{1}{|c|}{$k_K$} & \multicolumn{1}{|c|}{$k_P$} & \multicolumn{1}{|c|}{$i_{K}$} & \multicolumn{1}{|c|}{$i_{P}$ (deg)} \\  
\multicolumn{1}{|c|}{} & \multicolumn{1}{|c|}{$\pm$} & \multicolumn{1}{|c|}{$\pm$} & \multicolumn{1}{|c|}{} & \multicolumn{1}{|c|}{$\pm$} &
 \multicolumn{1}{|c|}{} & \multicolumn{1}{|c|}{$\pm$} & \multicolumn{1}{|c|}{(deg)} & \multicolumn{1}{|c|}{$\pm$} \\
\hline  \hline
1.01 & 1.00083 \,3 & \phantom{-}0.00001 \,2 & 0.118 & 0.122 \,2 & 0.124 & 0.1275 \,7 & 84.22 & 84.3 \,2 \\
\hline 
2.01 & 1.0044  \,1 & \phantom{-}0.000003 \,3 & 0.214 & 0.2155 \,3 & 0.075 & 0.0761 \,2 & 88.24 & 87.2\, 5  \\
\hline 
3.01 & 0.99400 \,1 & \phantom{-}0.00002 \,2 & 0.060 & 0.0891 \,2 & 0.058 & 0.0585 \,2 & 89.95 & 89.9\, --1\\
\hline 
4.01 & 1.00396 \,2 & \phantom{-}0.0005\, 2 & 0.223 & 0.2232 \,4 & 0.042 & 0.037 \,9 & 77.92 & 78 \,2 \\
\hline 
5.01 & 1.001 \,2 & -0.0001 \,2 & 0.132 & 0.133 \,1 & 0.037 & 0.036 \,1 & 82.51 & 82.8 \,5 \\
\hline 
7.01 & 1.0046 \,8 & \phantom{-}0.0005 \,3 & 0.223 & 0.2226 \,4 & 0.027 & 0.0269 \,7 & 80.79 & 80.6 \,6 \\
\hline 
10.01& 1.00137 \,2 & -0.0001 \,1 & 0.133 & 0.146 \,2 & 0.093 & 0.0972 \,8 & 85.37 & 84.0 \,4\\
\hline 
12.01 & 1.0001 \,1 & \phantom{-}0.00001 \,1 & 0.050 & 0.05420 \,6 & 0.088 & 0.0912 \,3 & 89.95 & 88.7 \,1 \\
\hline 
13.01 & 1.00401 \,8 & -0.00009 \,3 & 0.227 & 0.2315 \,2 & 0.078 & 0.0869 \,3 & 85.37 & 88.1 \,6 \\
\hline 
17.01 & 1.00112 \,7 & \phantom{-}0.00001 \,6 & 0.133 & 0.1366 \,1 & 0.094 & 0.0953 \,6 & 89.95 & 87.7 \,4 \\
\hline 
20.01 & 1.00084 \,6 & -0.00001 \,3 & 0.123 & 0.126 \,1& 0.117 & 0.1202 \,6 & 89.95 & 88.6 \,5 \\
\hline 
42.01(t) & 1.00053 \,1 & -0.00067\,3 & 0.052 & 0.053 \,1 & 0.018 & 0.0180 \,2 & 87.66 & 87.7 \,1 \\
\hline 
42.01(d) & 0.99875 \,1 & -0.00066 \,3 & 0.052 & 0.051 \,1 & 0.018 & 0.0176 \,2 & 87.66 & 87.7 \,1 \\
\hline
72.01 & 1.010 \,4 & \phantom{-}0.0006 \,8 & 0.280 & 0.2797 \,6 & 0.013 & 0.015 \,2 & 87.09 & 87 \,3 \\
\hline 
117.01 & 1.0010 \,2 & \phantom{-}0.00003 \,3 & 0.131 & 0.131\, 1 & 0.023 & 0.023 \,2 & 83.08 & 83 \,1 \\
\hline 
377.01 & 1.0004 \,7 & -0.00087 \,3 & 0.028 & 0.032 \,2 & 0.075 & 0.079 \,3 & 89.38 & 88.9 \,4 \\
\hline 
388.01 & 1.0010 \,1 & \phantom{-}0.00066 \,6 & 0.125 & 0.135 \,3 & 0.017 & 0.018 \,2 & 87.09 & 87.8 \,2  \\
\tableline 
\end{tabular} 
\end{table*}

\begin{table*}[]
\small
\caption{Additional fitting information}
\label{tab:no4}
\begin{tabular}{|r|r|r|r|r|r|l|r|r|r|}
\tableline 
\multicolumn{1}{|c|}{KOI} & \multicolumn{1}{|c|}{${d/{R_*}}_K$} & \multicolumn{1}{|c|}{${d/{R_*}}_P$} & \multicolumn{1}{|c|}{$b_K$} & \multicolumn{1}{|c|}{$b_P$} & \multicolumn{1}{|c|}{$u_{vH}$ } & \multicolumn{1}{|c|}{$u_P$} \\ 
 
\multicolumn{1}{|c|}{} & \multicolumn{1}{|c|}{} & \multicolumn{1}{|c|}{} & \multicolumn{1}{|c|}{} & \multicolumn{1}{|c|}{} & \multicolumn{1}{|c|}{} & \multicolumn{1}{|c|}{$\pm$} \\ 
\hline \hline

1.01 & 8.445 & 8.177 & 0.82 & 0.82 & 0.56 & 0.80 8 \\
\hline 
2.01 & 4.681 & 4.690 & 0.13 & 0.23 & 0.51 & 0.46 1 \\
\hline 
3.01 & 16.680 & 16.892 & 0.03 & 0.02 & 0.69 & 0.63 1 \\
\hline 
4.01 & 4.481 & 4.480 & 0.95 & 0.94 & 0.50 & 0.6 6 \\
\hline 
5.01 & 7.560 & 7.508 & 0.95 & 0.94 & 0.55 & --- \\
\hline 
7.01 & 4.486 & 4.492 & 0.71 & 0.74 & 0.54 & 0.54 2 \\
\hline 
10.01 & 7.500 & 6.873 & 0.64 & 0.72 & 0.52 & 0.59 9 \\
\hline 
12.01 & 20.040 & 18.450 & 0.03 & 0.43 & 0.50 & 0.35 1 \\
\hline 
13.01 & 4.400 & 4.170 & 0.35 & 0.47 & 0.44 & 0.51 1 \\
\hline 
17.01 & 7.530 & 7.321 & 0.03 & 0.29 & 0.55 & 0.53 3 \\
\hline 
20.01 & 8.133 & 7.937 & 0.02 & 0.19 & 0.52 & 0.48 2 \\
\hline 
42.01(t) & 19.400 & 19.015 & 0.77 & 0.78 & 0.51 & 0.5 3 \\
\hline  
42.01(d) & 19.400 & 19.473 & 0.77 & 0.77 & 0.51 & 0.54 9 \\
\hline
72.01 & 3.575 & 3.575 & 0.17 & 0.20 & 0.50 & 0.6 4 \\
\hline 
117.01 & 7.620 & 7.634 & 0.92 & 0.92 & 0.53 & 0.5 3 \\
\hline 
377.01 & 36.000 & 30.769 & 0.35 & 0.61 & 0.57 & --- \\
\hline 
388.01 & 8.020 & 7.446 & 0.39 & 0.29 & 0.56 & 0.3 3 \\
\tableline 
\end{tabular} 
\end{table*}

 We have worked with `normalized' light curves, for which the measured fluxes are divided by a representative mean out-of-eclipse value,  so that $U$ (`unit of light') is nominally unity, at least initially. $\Delta \phi_0$ is in units of the complete range 0-1.  The relative radius $r_1$  is the radius of star $R_1$ in units of the semi-major axis $a$,  $k$ is the ratio of planet to stellar radii ($R_p/R_*$), and $i$ denotes the inclination (in deg) of orbital axis to the line of sight.  For KOI 42.01 we list two sets of results: (t) the initial result, for which a linear trend across the transit phase range was noted (see Fig~24), and (d) after this trend had been removed from the input data.  This arrangement holds also for subsequent table entries for KOI 42.01.

As well as the main results in conventional notation for binary light curves, Table \ref{tab:no4} lists: ${d/R_*}_K$; the inverse of the star radius in units of the semi-major axis, as listed by the NEA, ${d/R_*}_P$ -- the same quantity from the present paper's fittings, $b_K$ -- impact parameter $=(R_{*}/a)\cos i$, as in the NEA, and $b_P$ as found by us. 
We also compare the stellar limb-darkening coefficient calculated by interpolating from the tables of Van Hamme (1993), $u_{vH}$, with the optimized value from the transit fitting, $u_P$. A dash indicates that a fully determinate solution  (positive definite error matrix at the optimum) was not found. In most cases, the optimized value is close to the van Hamme value. The abnormally high value found for KOI 1.01 may be a result of it having a (visual) binary companion (Daemgen et al, 2009),
thereby affecting the assigned (near solar) temperature.  

Table \ref{tab:no5} allows assessment of the quality of the light-curve fittings.  This involves $\chi^2/\nu$ which should be close to unity for a reasonably probable model fitting. $\Delta l_K$ is the NEA-derived data measure of datum accuracy, i.e.\ (mean error)/(mean flux), while $\Delta l_p$ is the Poisson factor given in Table \ref{tab:no2} divided by the scaling constant $s = 6.02$, which is the mean value of individual Poisson factor to $\Delta l_K$ ratios. The ratios $\Delta l_K/\Delta l_p$ are also listed. The  $\Delta l$s, whether from the NEA or the Poisson factors, scale the $\chi^2/\nu$ values to $\sim$ 1 reasonably well in most cases. There are anomalies in some instances, however.  The $\Delta l_K$ values take some account of variation in intrinsic scatter in the source counts at a given mean brightness (Gilliland et al.\ 2011), but if this is higher than normal in a particular case, an additional intrinsic variability, perhaps of a pulsational type, could be the explanation.  This seems feasible for KOI 10.01, where there is an unexpectedly high $\chi^2/\nu$ ratio, as well as a significant excess of  $\Delta l_K/\Delta l_p$.

The other examples of high  $\chi^2/\nu$ are KOI 2.01, 3.01, 7.01 and 42.01. In the cases of KOI 2.01 and 3.01, there are clear signs in the data of maculation effects: the eclipses of individual surface features being apparent in the light curves.  Such features are not taken account of in the modelling, so the excess of $\chi^2$ will reflect this easily identified extraneous effect.
For KOI 7.01, the scatter in the residuals curve shown in Figure 13, whilst apparently uniform,
appears noticeably wider than the 0.0004 value assigned.  This could then be another example
of micro-variability, although, in this case, the NEA error measure has not been appreciably
affected by the excess scatter from the candidate star.  

Something similar may apply also for the brighter KOI 42.01 (see Figure~24), though there are also longer term sources of flux variation in that case.  This example required us to deal with the issue of `cleaning' or detrending the data, which we discuss in more detail in subsection 3.11 below. 

\begin{table}[]
\caption{Quality of fit data}
\label{tab:no5}
\begin{tabular}{|r|r|r|r|r|c|}
\tableline 
\multicolumn{1}{|c|}{KOI} &  \multicolumn{1}{|c|}{$\nu$} & \multicolumn{1} {|c|}{$\chi^2/\nu$} & \multicolumn{1}{|c|}{$\Delta l_K$} & \multicolumn{1}{|c|}{$\Delta l_p$} & \multicolumn{1}{|c|}{ratio}\\
\multicolumn{1}{|c|}{} &  \multicolumn{1}{|c|}{} & \multicolumn{1} {|c|}{} & \multicolumn{1}{|c|}{} & \multicolumn{1}{|c|}{} & \multicolumn{1}{|c|}{}\\
\hline \hline
1.01 &  240 & 0.946 & 0.00026 & 0.00026 & 1.00 \\ 
\hline 
2.01 &  304 & 1.500 & 0.00015 & 0.00016 & 0.94 \\ 
\hline 
3.01 &  259 & 1.436 & 0.00009 & 0.00010 & 0.90 \\ 
\hline 
4.01 &  342 & 1.012 & 0.00026 & 0.00027 & 0.96 \\ 
\hline 
5.01 &  241 & 0.797 & 0.00028 & 0.00029 & 0.97 \\ 
\hline 
7.01 &  498 & 1.512 & 0.00040 & 0.00040 & 1.00 \\ 
\hline 
10.01 & 395 & 1.780 & 0.00097 & 0.00075 & 1.29 \\ 
\hline 
12.01 & 895 & 1.025 & 0.00026 & 0.00026 & 1.00 \\ 
\hline 
13.01 & 378 & 1.066 & 0.00011 & 0.00013 & 0.85 \\ 
\hline 
17.01 & 424 & 1.142 & 0.00062 & 0.00061 & 1.02 \\ 
\hline 
20.01 & 567 & 1.104 & 0.00077 & 0.00069 & 1.12 \\ 
\hline 
42.01(t) &  566 & 1.675 & 0.00009 & 0.00010 & 0.90 \\ 
\hline 
42.01(d) & 566 & 1.175 & 0.00009 & 0.00010 & 0.90 \\ 
\hline 
72.01 & 218 & 1.083 & 0.00022 & 0.00023 & 0.96 \\ 
\hline 
117.01 & 818 & 1.061 & 0.00044 & 0.00045 & 0.98 \\ 
\hline
377.01 & 506 & 0.935 & 0.00103 & 0.00085 & 1.21 \\ 
\hline 
388.01 & 672 & 1.083 & 0.00080 & 0.00068 & 1.18 \\ 
\tableline 
\end{tabular} 
\end{table}

\begin{table}[]
\small
\caption{Key output astrophysical parameters}
\label{tab:no6}
\begin{tabular}{|r|r|l|r|r|}
\tableline 
\multicolumn{1}{|c|}{KOI} & \multicolumn{1}{|c|}{$R_{* K}$} & \multicolumn{1}{|c|}{ $R_{* P}$} & \multicolumn{1}{|c|}{ $R_{p K}$ } & \multicolumn{1}{|c|}{$R_{p P}$} \\
\multicolumn{1}{|c|}{} & \multicolumn{1}{|c|}{($\odot$)} & \multicolumn{1}{|c|}{ ($\odot$)} & \multicolumn{1}{|c|}{ ($\oplus$) } & \multicolumn{1}{|c|}{($\oplus$)} \\    
\hline \hline
1.01 & 1.06 & 0.95 $\pm$ 0.02 & 14.40 & 13.17 $\pm$ 0.20  \\
\hline
2.01 & 2.71 & 1.81 $\pm $ 0.003 & 22.30 & 15.01 $\pm$ 0.03 \\
\hline
3.01 & 0.74 & 0.66 $\pm$ 0.002 & 4.68 & 4.23 $\pm$ 0.02 \\
\hline
4.01 & 2.60 & 2.69 $\pm$ 0.005 & 11.80 & 10.70 $\pm$ 0.20 \\
\hline
5.01 & 1.42 & 1.66 $\pm$ 0.01 & 5.66 & 6.45 $\pm$ 0.10 \\
\hline
7.01 & 1.27 & 2.11 $\pm$ 0.004 & 3.72 & 6.18 $\pm$ 0.20 \\
\hline
10.01 & 1.56 & 1.47 $\pm$ 0.02 & 15.90 & 15.60 $\pm$ 0.20 \\
\hline
12.01 & 1.40 & 1.68 $\pm$ 0.002 & 13.40 & 16.70 $\pm$ 0.20 \\ 
\hline
13.01 & 2.70 & 1.84 $\pm$ 0.002 & 23.00 & 17.46 $\pm$ 0.03 \\
\hline
17.01 & 1.08 & 1.32 $\pm$ 0.001 & 11.04 & 13.74 $\pm$ 0.04 \\
\hline
20.01 & 1.38 & 1.52 $\pm$ 0.01 & 17.60 & 19.90 $\pm$ 0.20 \\ 
\hline
42.01(d) & 1.36 & 1.59 $\pm$ 0.05 & 2.71 & 3.06 $\pm$ 0.10 \\
\hline
72.01 & 1.00 & 1.02 $\pm$ 0.002 & 1.38 & 1.64 $\pm$ 0.01 \\
\hline
117.01 & 1.18 & 3.52 $\pm$ 0.03 & 2.93 & 10.19 $\pm$ 0.40 \\
\hline
377.01 & 1.01 & 1.00 $\pm$ 0.060 & 8.28 & 8.65 $\pm$ 0.50 \\
\hline
388.01 & 1.58 & 1.96 $\pm$ 0.040 & 3.01 & 3.75 $\pm$ 0.40 \\
\tableline
\end{tabular} 
\end{table}

In Table \ref{tab:no6}, the absolute radii of host stars ($R_*$) are given in solar units and those of the  planets ($R_p$) in units of that of the Earth. In the second column, the host star radii are the NEA values using the equation $R_*$ = $\sqrt{M_*g_\odot/g_*}$. In column 3 the radii $R_{*P}$ are derived from multiplying the fitted $r_1$ values by the NEA value of the semi-major axis, $a$. The difference between the two $R_*$ estimates can be seen to be quite appreciable in some cases, with the larger range of values generally corresponding to the  $R_{*K}$ numbers, as previously suggested. The exceptional case of KOI 117 is discussed in Section 3.13, below. The planetary radii are obtained by multiplying the stellar radii by the ratio of radii, i.e. $R_{p K}$ = $R_{*K}k_K$ and $R_{p P}$ = $R_{* P}k_P$. For KOI 42.01 we retain only the final detrended results, that have significantly reduced $\chi^2/\nu$ values.

\section{Individual systems}
In what follows we give summary notes on each of the examples studied.
  
\subsection{KOI 1.01}
 KOI 1.01, also known as TrES-2b and Kepler-1b, was identified as a transiting exoplanet well before the Kepler Mission (Sozetti et al, 2007). The star system has been identified as a binary consisting of a G0 star, around which the planet orbits, and a K4.5-K6 red dwarf (Daemgen et all, 2009). It appears to be a relatively uncomplicated light curve, in which a `hot jupiter' transits the parent star towards its limb. The light curve data for KOI 1.01 consisted of the section from BJD 122.68 to 122.84 of the short cadence data set from quarter 0. An initial solution using only the first few transits seemed better fitted by a partial transit ($\cos i > r_1 - r_2$, where $r_2 = r_1\* k$), but the rounded bottom of the eclipse becomes clear with sufficient coverage. The result is a complete eclipse, although the transit is quite near to the limb. 

The NEA gives the relative radius of the star $r_1$ as 0.118, compared to our 0.1223$\pm$0.0007. The physical sizes are: $R_{* P}$ = 0.95 $R_{\odot}$, and $R_{p P}$ = 13.11 $R_{\oplus}$. This is in fair agreement with the NEA listing of 1.06 $R_{\odot}$ and 14.40 R$_{\oplus}$ respectively.

\subsection{KOI 2.01}
 KOI 2.01 (= HAT-P-7, Kepler-2) has become a well-known exoplanet system since its discovery by Pal et al.\ (2008).  The relatively large star in this system with its close-in planet makes for a fairly wide transit of duration around 10\% of the complete phase cycle. Our light curve for KOI 2.01 comes from section JD 121.25-121.46 of the short cadence data of quarter 0. The NEA value for  $r_1$ is 0.214 is in essential agreement with our value of 0.2155$\pm$0.0003. But we find  a significantly lower value of the  inclination, 87.2$\pm$0.5 deg versus the NEA value of 88.24 deg. Our ratio of radii, 0.0761, is not far from that of the NEA value, 0.075, so our solution would yield a fairly large planet, 14.40 $R_{\oplus}$ which is significantly smaller than the NEA value of 22.30 $R_{\oplus}$.

The transit minimum shows clear signs of a spot eclipse. This lasts for about 8 deg of phase during the ingress, or about 0.14 of the stellar diameter, i.e.\ this spot seems likely to have a significantly larger size than the planet ($\sim$4.5 $R_J$). It must therefore be much larger than known solar spots. The modelling of spot eclipses considered by Budding (1988) could not be used directly here, since that applied only to the occultations of spots by relatively large bodies. 

KOI 2.01's relatively massive planet is close in to the star in a configuration comparable to that of KOI 13.01, for which proximity effects are measurable from the light curve (see below). We therefore sought to fit the full light curve for this system using the Radau model and the `black body' approximation for the light redistribution (Budding \& Demircan, 2007; p319) to scale trial proximity effects.  These effects appear masked by the maculation effect and perhaps other contributions to the light curve segments we studied, rendering our proximity term evaluation unreliable.  However, Welsh et al (2010), after detailed processing of 15 short cadence light curves, derived a mass ratio of about 0.013 using the Roche model in this way.

\subsection{KOI 3.01}

 KOI 3.01 is another object whose planetary status was studied by the earlier HATNet Project  (Bakos et al, 2002, 2009, 2010). The star is also known as HAT-P-11 and, as a confirmed planet, Kepler-3. The light curve of KOI 3.01, shown in Fig 3, and taken from BJD 144.27-144.45 of the short cadence data of quarter 1, corresponds to the wider separation and smaller star and planet than for the first two examples. The star is somewhat brighter than in KOI 2.01, but it is a lower mass object. Again there are signs of surface inhomogeneities, this time relatively small in size. Our finding that the planet is about 4.23 times bigger than the Earth in radius agrees closely with the NEA value of 4.68 R$_{\oplus}$, and in apparent exact agreement with that of Bakos et al.\ (2010).

Bakos et al (2009, 2010) found an eccentricity of 0.198 to the planetary orbit, having a periastron longitude of {$\sim$355\degr} with a formal error of about {15\degr}.  In other words, the major axis of the elliptical orbit is not far from the plane of the sky, the planet exiting from the transit more quickly than its entrance. We have checked this in our fitting program.  The transit's fitting is improved by taking its possible asymmetry into account, and the determined value of the optimal periastron longitude corresponding to the assigned eccentricity $e$ is about {8\degr}.  The determination is relatively good, if we suppose all the other parameters are already well-defined, because the degree of asymmetry of the transit is close to the maximum possible for the assigned value of $e$.  However, this result should probably not be taken at face value.  The improvement in the transit fitting can be associated with a better match of the model of light variation through the region of the eclipsed spot. Disentangling the photometric effects of maculation from those of orbital eccentricity is not straightforward for the empirical analysis of a single transit, as we are doing. Bakos et al.\ (2009) also noted the complications to parameter estimation arising from the interactions of such varied causes of data variation. For the present, therefore, this apparent confirmation of the orbital eccentricity should be regarded cautiously.

\subsection{KOI 4.01}

We took the light curve for KOI 4.01 from section BJD 172.80-173.04 of the short cadence data in quarter 2. With an transit depth about 1/4 that of the first few examples, KOI 4.01, a fainter source than the others so far, exhibits a much less well-defined light curve. The derived planet size turns out to be quite comparable to that in KOI 3.01, while the star itself is appreciably bigger, so the transit effect is smaller. There is nothing obvious in these data that would immediately indicate a 'false-positive', but KOI 4 was given this status, apparently due to inclusion of a background eclipsing binary in the photometry. This points up the possibility that other {\em prima facie} planetary light curves turn out to be false positives. 

Here we note that a false positive occurs when the star identified with a planetary transit is not so eclipsed.  There may still be a planetary transit within the photometric aperture, but small shifts in the light centre during the eclipse indicate the effects relate to another star, not the one identified in the catalogue (see Bryson et al, 2013, for a recent review).
In fact, a plausible alternative solution was found, corresponding to a background Algol system about 6.4 mag fainter than the target star. The small relative radius of the eclipsed star, required by the low phase range of the eclipse, combined with the 3.85 d period, would require the background system to be relatively massive -- perhaps $\sim$5 solar masses or so, but the resulting $\chi^2/\nu$ value (1.07) is hardly different from the planetary transit model in terms of probability.

There is a distinct maculation effect seen in the out-of-transit light of this star. Periodogram analysis (available as an on-screen option on the NEA data display webpage) indicated a likely value of $\sim$5.8 days for the stellar rotation period as a result.

\subsection{KOI 5.01}

 In the case of KOI 5.01, there was a reasonable agreement between our starting trial value of $r_1$, using the $R$ and $a$ values of Table \ref{tab:no1}, and the value obtained from the curve-fitting. It is thus not surprising that our final value of $R_p$, at 6.45 R$_{\oplus}$, is not far from the NEA value (5.66 R$_{\oplus}$). The transit is complete, although close to the limb. The data used for this candidate are from section BJD 171.13-171.30 of the short cadence measures from quarter 2.


\subsection{KOI 7.01}

 KOI 7.01, also known as Kepler-4, was reviewed by Borucki et al.\ (2011a). With KOI 7.01 we again find a relatively small planet in a not-too-distant orbit. Although the loss of light is similar to KOI 4.01, the flux is only 1/3 of that, so the low S/N of the light curve makes for a generally imprecise fitting. Nevertheless, our value for the planetary radius, 6.18 $R_{\oplus}$, is significantly larger than the NEA value of 3.72 $R_{\oplus}$, and is also larger than than the 3.99 $R_{\oplus}$ found by Borucki et al.\ (2011a).

As indicated in the notes on Table 5 in the previous section, the scatter of the data in the uniform distribution of residuals is appreciably larger than the NEA numbers suggest and micro-variability of some kind may be the explanation. We followed the option of carrying out a periodogram analysis for this data-set, which comes from section BJD 171.65-171.99 of the short cadence run in quarter 2.  We found the power spectrum having spectral components of much lower frequency than those corresponding to small pulsations. The present paper, whilst leading to suggestions about candidate micro-variables, does not address the detailed properties of such stars.

\subsection{ KOI 10.01}

 The flux for KOI 10.01 is an order of magnitude less than for KOI 1.01, though the depth of transit is comparable in this data section from BJD 173.82-174.09 of the short cadence, quarter 2 run.
 We therefore do not expect so good a solution for this large planet, bigger than Jupiter by $\sim$40\%. The scatter in the solution is $\sim$5 times bigger than that of KOI 1.01. As with KOI 7.01, the star may therefore be showing some additional effect that the modelling does not take into account, perhaps small-scale, short-period inherent fluctuations. The relatively large value of $\chi^2/\nu$ in Table \ref{tab:no5} supports the same inference.
 
\subsection{KOI 12.01}

 The planet in this system has the greatest separation from its host star in our sample, at just over 30 solar radii, but its estimated mean surface temperature at over 800 K is still too high for the `habitable zone'.  The star is not bright, so the combination of short transit phase range and scatter makes for a relatively poor solution. The preset value of the limb-darkening coefficient given in Table \ref{tab:no1} (0.50), produced a systematic trend in the residuals. Our optimized value (0.35) has effectively removed this irregularity. Since optimized limb-darkening coefficients, when allowed to be free parameters, generally agree with those coming from the van Hamme (1993) code, this disagreement is suggestive that one of the underlying parameters for the code, probably the star's effective temperature, has been wrongly assigned in the NEA.
The used light curve data are from BJD 128.43-129.05 of the short cadence set from quarter 0.

\subsection{KOI 13.01}

 KOI 13.01 has become a well-known candidate of special interest, since it was noticeable, even from the preliminary inspections, that the light curve shows a secondary minimum and proximity effects. This accords with reasonable expectation for a relatively large planet at fairly low separation. Photometry of this object is, however, compromised by the presence of a close companion of comparable brightness, which entails slight shifts of the light centre on the detection array that may affect the processing statistics. Some data-sets also show indications of additional light variations, perhaps related to maculation effects, or the role of the additional low mass companion in the host star multiple system recently identifed by Santerne et al.\ (2012).  Welsh et al.\ (2010) mentioned also the possibility of a focus drift.
 
Although the transit fitting, that corresponds to data selected from BJD 169.81-170.07 of the short cadence second quarter run,
 may look good, judging from a fairly uniform scatter of residuals, such difference curves have sometimes shown residual short-term irregularities (blips) at the inner contact points.  In this connection, we anticipated a possible failure of the limb-darkening approximation very close to the limb. A second order limb-darkening term was therefore included in the fitting function.  However, it became clear that this enhancement, whilst slightly lowering the optimal $\chi^2$ value, did nothing to remove the blips. Later experiments have shown that such effects are associated with a poor selection of the orbital inclination.  This controls the rate at which the planet passes over the limb, and that plays a critical role for the blips. The presence of blips in an otherwise satisfactory set of residuals lends a good diagnostic quality about the action of the optimization, and for the inclination in particular. Removal of the end-point blips in fitting the transit gave a definitive quality to the corresponding main geometrical elements.  Figure \ref{fig:KOI13} shows the fitting to the primary transit used to derive the parameters given in Tables \ref{tab:no3} and \ref{tab:no4}.

There also appears a slight residual trend both through the transit region and a little way on either side. This may be associated with the flux centroid drifting, mentioned above (see also Mislis \& Hodgkin, 2012). On the whole, we regard our modelling of this complex system hitherto as preliminary. The object will merit much closer individual attention. 

As mentioned, KOI 13.01 is associated with normal proximity effects throughout its orbital cycle. To examine these effects we used normalized long cadence data from quarter 2 extending from BJD 208.507 to 210.612. Figure \ref{fig:KOI13ecl} shows the out-of-transit region of the light curve of KOI 13 on an expanded scale. The primary minimum region has been clipped. The secondary eclipse of the reflected light is visible, as are the proximity effects. The first maximum is seen to be slightly higher than the second one, in agreement with the effect of `Doppler beaming'.  Doppler beaming is programmed to correct the assigned relative luminosity of the host star $L_1$ (essentially unity, unless another star is contributing to the measured flux), so that instead of $L_1$ in the fitting function (Budding \& Demircan, 2007; Eqn 9.17)  we use $L_{D1}$, say, through the formula
\begin{equation}
 L_{D1} = L_1(1 + 2v_{z1}/c) \,\,\, .
\end{equation} 
Here the line of sight component of the (stellar) orbital velocity is given by $v_{z1} = q v_{\rm orb} \sin \phi \sin i /(1+q)$, $q$ being the mass ratio planet to star, $\phi$ the orbital phase and $i$ the inclination. The velocity of light $c$ is set at 299792 s$^{-1}$. The orbital velocity is determined from known parameters as $v_{\rm orb} = 2\pi a/P$. Using the value of the planetary mass from Table \ref{tab:no1} the effect turns out to improve the fitting, but only at a marginal level of significance: $\chi^2$ decreasing by an amount of order unity  after the replacement of $L_1$ by $L_{D1}$ in the fitting function. 

As noted by Mislis \& Hodgkin (2012), the proximity effects offer a way to determine the mass ratio, although various other parameters are involved in fixing the scale of these effects. One parameter is $r_1$, which scales the `ellipticity effect' with a third power dependence --- so clearly having relatively high sensitivity. We found the out-of-eclipse variation tending to bias $r_1$ to a slightly larger number than that corresponding to the optimal fitting of the transit alone.  However, since the transit phases are relatively clear of proximity effects, we felt it legitimate to use the $r_1$ determined from the transit to throw the weight of the complete light curve fitting onto the mass ratio. Actually, the mass ratio ($q$) and gravity-brightening coefficient ($\tau$ --- another
unknown parameter producing the out-of-eclipse `ellipticity effect' light variation) are both involved in the same linear combination, so if either is decreased the other must be
correspondingly increased to produce essentially the same result. The structural parameters also influence the scale of the ellipticity effect in the same kind of linearly correlated way as $\tau$, so their inclusion tends simply to scale down the value of $q$ in proportion to the increase in the 
$ \Delta$ coefficients (see Introduction section) above unity. 

We  adopted the value of $\tau$ according to its `black-body' formula given in {\sc WinKepler}, allowing the ellipicity effect to determine only $q$. The value of $q$ found in this way (0.0033$\pm$0.0005) compares well with that (0.0029$\pm$   0.0005) given by Mislis \& Hodgkin (2012).  We should not regard this close numerical agreement as definitive, however.  As we have seen the light curves of KOI 13.01 are complicated by other effects not included in our present model, while even within that model there is scope for the variation of a combination of other parameters that can mimic the effect of a given mass ratio. Hence, the formal error on $q$ is really only a lower limit, corresponding  to what could be estimated as the error if all the other parameters were regarded as precisely known.  Still, the general consistency about  the result is encouraging (see also Welsh et al., 2010; Mazeh et al., 2012).
  
\subsection{KOI 17.01}

The Jupiter-sized planet, KOI 17.01, produces a clear and relatively deep primary minimum, even if over a small phase range. The photometry, corresponding to BJD 173.09-173.39 of the short cadence second quarter run, shows a rather higher level of scatter in the fitting than might have been anticipated (cf.\ Figs~21 and 22). There may then be some residual variation in the light curve, though not as pronounced as for some other examples.  

\subsection{KOI 20.01}
Although a somewhat fainter and noisier data-set than that of KOI 17.01, the light curve shows a relatively deep eclipse (in fact the deepest in our selection) caused by the  near-central transit of large ($\sim$1.6 R$_J$) planet in a comparable configuration. Our light curve for KOI 20.01 came from from BJD 170.81-171.20 of the short cadence data set from quarter 2.

\subsection{KOI 42.01}
 This planet with radius at about 2.7 R$_{\oplus}$, is one of the smaller ones in our list. Taking into account also the wide separation of close to 30 solar radii, the determinacy of the transit is not high. The flux level is relatively high, however, so a solution is possible in which the planet passes not far from the limb of the host star. The fitting is shown in Figure \ref{fig:KOI42}. The data were taken from BJD 145.36-145.75 of the short cadence first quarter run.

The residuals curve in Fig~24 shows a small downward trend, and an additional variability for this star was mentioned in Section 2.2 in connection with the anomalously high value of $\chi^2/\nu$.  Following the approach  of Zeilik et al (1988) in dealing with light curves affected by starspots,
 a `cleaning' stage after the first fitting was considered.  The residuals curve may be fitted by a linear regression,  the original transit data then `corrected' to a `clean' input file,  and the fitting run again.  The main geometric parameters  all have an even-function effect in varying the light curve about zero phase though, so, while the residuals would be reduced by such a correction and the $\chi^2/\nu$ anomaly reduced, the optimal parameter-set of the model should not be greatly affected. The downward trend across the range is not more than $\sim$20\% of the mean scatter, so we would expect the maximum effect on the $\sim$70\% excess of $\chi^2/\nu$ in the original fitting to be not more than $\sim$40\%.  The initial impression from Fig~24 is that random scatter is greater than the assigned 0.0001, and a small pulsational behaviour can be anticipated. 
 
 In fact, the fitting to the detrended data-set did reduce $\chi^2/\nu$ by most of its original excess, while there were also small changes to the main parameters, particularly $r_1$. It is known that the optimization convergence is slowed by the presence of extraneous effects, so it can be presumed that the detrending accelerated the fitting process to an inherently better set of parameter estimates. The $\sim$17\% remaining excess of $\chi^2/\nu$ may still reflect some inherent short-period variability.

\subsection{KOI 72.01}

This system has the shortest period of our sample, and the planet's radius is found to be only about 1.64 R$_{\oplus}$, making the planet a `hot Earth', orbiting close in to a sunlike star. The transit is relatively wide in phase, therefore, but not in time. Figure \ref{fig:KOI72} shows the fast ingress and egress, with data points evenly spaced. The data are from BJD 263.82-263.97 of the short cadence observations in the second quarter.

The transit, which appears shallow against the scatter, is not well resolved in this way. Even with this non-optimal monitoring arrangement, however, Figures~25 \& 26 show that an acceptable  solution is possible.

\subsection{KOI 117.01}
 A shallow and narrow eclipse makes for a relatively poorly defined parameter-set for this rather faint system that resembles KOI 42.01. Our light curve is from BJD 271.24-271.80 of the short cadence data set from quarter 2. The transit appears free of additional complications and the $\chi^2/\nu$ ratio and distribution of residuals imply the main photometric elements, which are not so different from those of the NEA data-base, are reliable.   Yet the originally assigned NEA radius of the host star (from the gravity formula) is appreciably different from the value found by our program.  This latter radius is relatively large: at around 3 solar radii it is the largest star in the sample, and presumably corresponds to a relatively old object now well on towards the end of its Main Sequence state. 

\subsection{KOI 377.01}

 KOI 377.01 is also known as Kepler 9-b and formed part of a detailed review  by Torres et al.\ (2011). KOI 377 contains at least one similar, but more distant, planet as well as probably a third inner and smaller one. Torres et al.\ (2011), concerned with the occurrence of false positives, or non-planetary configurations that might simulate similar transit-like effects, were able to show with confidence that KOI 377.01 is a Saturn-sized planet.  Time of transit variations in this
system were previously studied by Holman et al.\ (2010).

The data used for KOI 377.01 came from BJD 644.89-645.24 of the short cadence run from quarter 7,
from which we can clearly see that this solar-type star is affected by significant ($\sim$0.5\% of the mean flux) out-of-transit variability, probably maculation. There are also a couple of rather high `rogue points'.  Such occasional discordant individual points have been noticed in the data-sets examined for this study. They may perhaps result from cosmic ray effects not removed in preliminary processing of the archived data. Or, in keeping with the idea of enhanced magnetic activity associated with strong maculation, one could perhaps also expect occasional flares. Rogue points tend to slow convergence in the modelling and it is advisable to remove them when possible.

The regular approach to maculation effects has been to detrend beforehand (cf. KOI 42.01 above), although the phase range of the transit is often so small as make the maculation variation of very little effect during a single transit, as here, particularly as the light loss is relatively deep for this central transit. Figure \ref{fig:KOI377} shows an 8 deg phase range about the transit displaying a good fit for the model. Periodogram fitting to the out-of-transit variation indicated a $\sim$8.4 d period, which can be associated with the stellar rotation. 

The planet, with less than 700 K mean temperature, is the coolest in our sample, but still quite far from being a `habitable zone' candidate.

A significant eccentricity was given for this system in the (unsourced) NASA Ames Research Center website (http://kepler.nasa.gov/Mission/discoveries/). As with KOI 3.01, we tested the listed value, $e$ = 0.15, with a transit fitting, having first produced a trial value for the longitude parameter $\omega$. In practice, it is the mean anomaly at phase zero  $M_0$ which requires specification for an eccentric light curve fitting, but $\omega$ is then determined by the formula $\omega =$ 90\degr $- \nu(e,M_0)$ (Budding \& Demircan, 2007, p314).  However, our transit fitting experiments were not decisive on this.  A wide range of values of $M_0$ seemed to give almost the same final $\chi^2$ value, while if we allow also the zero phase parameter $\Delta \phi_0$ also to be simultaneously adjustable the solution loses determinacy.   Basically, the scatter is too large in this data-set for a definitive assessment of the system's possible eccentricity from transit fitting alone, and even if numbers are produced for the eccentricity parameters they must be compromised by the aforementioned additional light variation effects.  The planet's longitude of periastron $\omega$ = 269.4 $\degr $  corresponding to the marginally lowest value of $\chi^2$, has the major axis of the ellipse close to the line of sight, i.e.\ an almost symmetrical transit.  This point by itself argues for poor determinacy of the eccentricity parameters from transit fitting alone, in this case.

\subsection{KOI 388.01}
 Figure \ref{fig:KOI388} shows the transit of a relatively small planet (`super-Earth') about one of the lower mass stars in our sample, that has a relatively low flux level. The transit is thus very noisy, like KOI 72.01 but worse, yet a 4-parameter determinate solution (the relatively independent phase correction $\Delta \phi_0$ having been eliminated in a preliminary run) was found. High determinacy for such a data-set cannot be expected, but it is worth considering in the context of small planet discoveries that the Kepler Mission addresses.  

The light curve data for KOI 388.01 was drawn from BJD 741.22-741.68 of the short cadence data in quarter 8.

\section{Summary and conclusions}

\subsection{The use of {\sc WinKepler}}

We have presented an independent approach to the analysis of archival photometric data from the Kepler Mission, applying alternative curve-fitting techniques with a modified version of the {\sc CurveFit} program for close binary systems (Zeilik et al., 1988): {\sc WinKepler}. The fitting function  involves tidal and rotational distortions of the components that depend on structural coefficients derived for modern stellar models. This permits, in principle, checks on stellar structure or the role of rotation that is non-synchronized to the planet's orbital motion. The important issue of determinacy for the underlying model is also checked in {\sc WinKepler} by examination of the curvature Hessian at the optimum. 

Several results of interest have emerged. For example, the data on KOI~7.01, 10.01 and 42.01 point to additional microvariability, apart from the main transit phenomena analysed. Similarly, the light curves of KOI 2.01 and 3.01 show clear evidence of surface inhomogeneities.  Such effects, not included in the present model, are reflected in high optimal values of $\chi^2/\nu$. The case of KOI 42.01 raised the issue of detrending. Since we dealt mainly with transits that often correspond to very narrow intervals of phase, the fitting function's inherent symmetry across the eclipse would often not be compromised by a slight linear gradient in the observations. For KOI 42.01 the value of $\chi^2/\nu$ was significantly affected, however, and linear detrending before the final fitting was advantageous for a more rapid convergence towards a realistic $\chi^2/\nu$ value and parameter-set.  The difference in the resulting geometric elements after the detrending was quite small in comparison to the derived errors, however.

We have shown that in most cases there is sufficient information in the transit sections of the light curves alone to allow a 5, often 6, element specification of the model. More information would be contained in light curves having a complete phase range, but for many of these sunlike stars there are relatively strong maculation effects that are separate from the parameter-sets of direct relevance to the present study and complicate
the out-of-eclipse variations on a relatively large scale.  In a few cases, we could estimate a likely rotation period using the maculation effect.

For KOI~4.01 we found a solution that was close to that previously published in the NEA, and essentially comparable to all the other normal star and planet combinations in this paper. However, Matijevic et al.\ (2012) cast doubt on this candidate as an exoplanet, indicating a more likely background close binary star. If such an ambiguity can exist among such normal-seeming candidates the implications for discoveries of exoplanets by the transit method could be seriously compromised, unless checks for background eclipses associated with active pixel offsets are simultaneously available (Bryson et al., 2013).  In that case, it may be possible to confirm the real source of the variation ascribed to a planetary transit of a given star.  Bryson et al.\ (2013) show that the planetary transit model for such a light curve may give an obviously poor fitting result, that would
be signalled by the $\chi^2/\nu$ value at the optimum. It is also still possible for the eclipse to
result from a planet, but not of the target star. The situation can be checked, in the curve-fitting context, by relaxing the condition that the fractional light of the eclipsed star is effectively unity, as in the normal case. For KOI~4.01, we thus found that the light curve could be modelled by a background classical Algol as effectively as by a planetary transit.  In the absence of independent
evidence coming from apparent shifts of the light centre on the detector array, ambiguity on the cause of the eclipses would remain.

Checks of theoretical linear limb-darkening coefficients by perturbing initially set values and allowing subsequent optimization have shown the values of van Hamme (1993) to be generally supported to within reasonable accuracy limits, though this was not the case for KOI 1.01 and 12.01. The assigned temperatures may be inappropriate for some reason, or the effects of interstellar extinction considered, in such examples.
The inclusion of a second order term in the limb-darkening approximation had only a marginal effect on  $\chi^2/\nu$: the expected amplitude of this term would  normally only be of the order of accuracy to which the linear term can be specified.

KOI 13.01 is noticeably different in several respects. Here, we were able to test the program, including its newly added Doppler beaming component, to the complete light curve. The significance of this latter component was demonstrated by a small reduction of the fitting's $\chi^2$. The effects of structural coefficients in the fitting function were also checked for this example.  Since the proximity effects scale mainly as a direct product of $q$ and $w_2$ a few percent decrease in $q$ is observed in moving from the Roche ($w_2 = 1$) to Radau ($w_2 \approx 1.05$) models. We expect the significance of this point to become enhanced in future cases of well-defined light curves containing large (Jupiter sized or greater) planets of short period, whose mass ratios may be independently checked spectroscopically.

\subsection{Comparison with the NEA parameters}
In Figure \ref{fig:RatioOfR_1} we show the trend of our results for the relative radius $r_{1P}$ against those published by the NEA $r_{1K}$. The regression coefficient (at 0.9978) looks reassuring, although we may notice that the $r_{1P}$ values of the present paper are generally slightly greater than those of the data-base $r_{1K}$ by a few percent. 
The recent addition of error estimates to the NEA parameter listings, together with those of the foregoing tables,
allow a better perspective on the generally low significance of
the differences in $r_1$ values.
The situation is similar with the ratio of radii (Fig \ref{fig:RatioOfRadii}), apart from KOI 13, where we find a distinctly smaller planet than that of the NEA.  Although, for some reason,
the NEA error estimates for $k$ appear about an order of magnitude lower
than our (interdependent) ones, the discrepancy for KOI 13.01 is outside
the range of assigned errors, unlike KOI 4.01 and 10.01.

It is with the distribution of inclinations (Fig \ref{fig:RatioOfInclination}) that we find a small but significantly different  trend, in the sense that the NEA inclinations are peaked more towards the 90\degr\ limit than ours.  Running a check for a random distribution of detectable (totally) eclipsing binaries, i.e.\ satisfying $r_1 - r_2 > \cos i$ for a given value of $r_1$, demonstrates the expected flatness of the distribution for high values of $i$. It is well known that, in general, the distribution of observed inclinations of binary orbits on the sky is proportional to $\sin i$. This means that the distribution of observed inclinations can be reproduced from  a uniform distribution of numbers in the range 0 to 1 by considering $\arccos i$ for those numbers.   Alternatively, the tendency to constancy of the parent distribution $\sin i$, when $i$ is close to 90 degrees, implies that  the numbers in each degree interval close to 90, for a given representative value of $r_1$, should be approximately similar. If a significant number of long period planets are included, the observed inclinations would compact towards 90, due to the decreasing value of $r_1$, but these cases are less frequently detected, and we have  only 3 planets with a period longer than 10 d.  Planets of solar type stars that cluster around an average 3 d period should have their inclinations distributed reasonably uniformly over the range $\sim$82-90 degrees. 

We have found, in fitting procedures that start with a value of $r_1$ that is too small, that most of the transit can be well fitted with a high value of $i$. Residual small blips at the ingress and egress points in the difference curve may remain, however, even though their effect on the net value of $\chi^2$ is small.  This possible `$\chi^2$-valley' in the optimization process may have occurred for NEA solutions whose inclinations appear somewhat clustered towards 90 degrees.

\subsection{Final remarks}

Our independent testing of the parametrization of light curves from the NEA shows a good measure of agreement in most cases. Where differences were found, we have inquired into their causes. We believe some differences arise from the more physically detailed fitting function and thorough optimization sequences of our findings. It is possible that local minima for $\chi^2$  occur for certain parameter combinations. The issue shows up in the difference (residuals) curve as fairly uniform scatter over most of the range, but sometimes with a few short-term irregularities, typically in the ingress and egress regions. Further optimization experiments with small random shifts in initial parameter sets near known approximate solutions can remove these.  

Some of the differences in our absolute parameters may come from a different procedure for finding the host star radius. The NEA appears to have retained $R_{*}$ values using the gravity estimates, as mentioned in Section 2.1. Here we can keep in mind the relative accuracies and sensitivities.  Semi-major axes calculated using the very precisely known period may have a reasonable accuracy despite the imprecisely known host star mass,
since the sensitivity of $a$ to $M$ is low. On the other hand, the mean density estimate is sensitive to $r_1$ and the gravity cannot be determined photometrically with sufficient precision at present.  

Our values of $\chi^2/\nu$ adopted as optimal are mostly consistent with a high probability to the underlying (eclipsing binary system) model. This is on the basis of error estimates for the data-sets supplied directly by the NEA and checked by reasonable statistical arguments on the flux detection. A few fittings result in  $\chi^2/\nu$ values that are  unusually high, pointing to the occurrence of inherent microvariability, surface maculation or both.

The possibility of empirical checks of limb-darkening, gravity-brightening, structural and rotational effects (in selected examples), gave additional objectives to the present work.  
Only KOI 13.01, in our sample, shows a sufficient scale of proximity effects (above other causes of flux variation) to allow testing of the relevant parameters (gravity darkening, reflection and Doppler beaming). The  mass ratio could be optimized and the empirical result is within reasonable error limits of other determinations of this value, although there are
still reasons to be cautious about the value of $q$ for the complex system KOI 13.01 at present.

A main argument in this article has been that it is useful to compare different analytical approaches. We have presented the new, purpose-designed, software {\sc WinKepler} to this aim.
This will help to make finally accepted stellar and planetary parameters and their corresponding error estimates, as obtained from transit photometry, more robust and better understood. Our initial sample of 16 planetary candidate light curves is perhaps rather small, but a methodologically useful avenue into the special issues raised by planet eclipses. We plan to continue with more detailed studies of individual data-sets using this software.

\acknowledgements 
 
MR and EB attended the Carl Sagan Summer Workshop at the California Institute of Technology, Pasadena during the week July 22-29, 2012, which was a considerable stimulation for much of the work in this paper. The meeting was organized by the NASA Exoplanet Science Institute (NExScI), who supported EB's  registration and accommodation.  EB appreciates also the  RASNZ's Kingdon Tomlinson Fund for its partial support of his travel costs.

%
%

%
%

%

%
%

%


%

%

\section{Appendix: Graphs of Fittings}

\textbf{See Figs. 2-39.}

\begin{figure}[H]
\begin{center}
\includegraphics[width=\columnwidth]{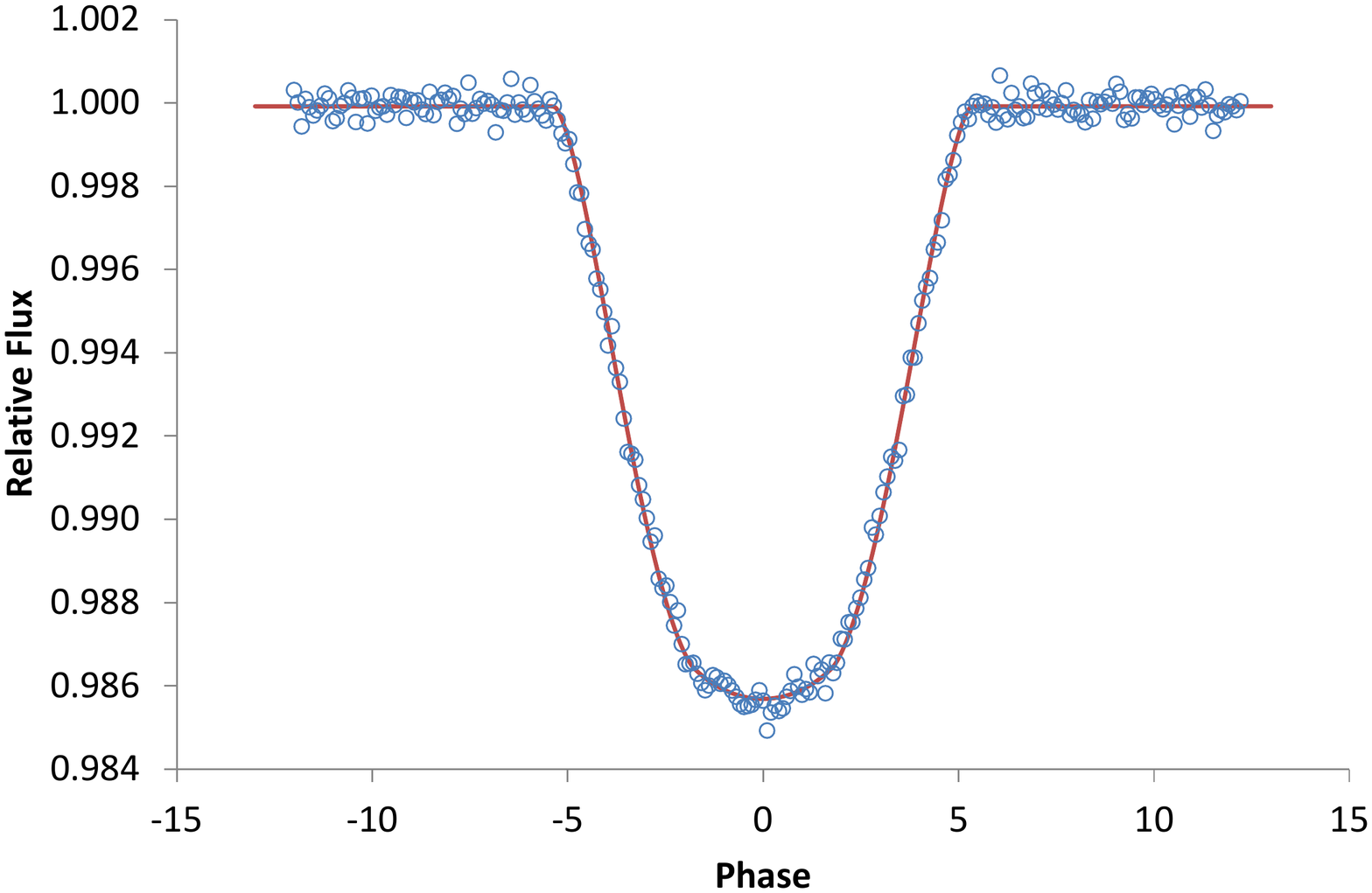} 
\caption{NEA light curve of KOI 1.01, matched by the {\sc WinKepler} model}
\label{fig:KOI1}
\end{center}
\end{figure}

\begin{figure}[H]
\begin{center}
\includegraphics[width=\columnwidth]{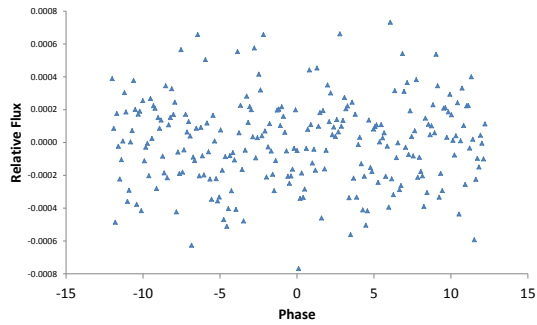} 
\caption{Difference between observed and calculated (residuals)  points for KOI 1.01.}
\label{fig:KOI1d}
\end{center}
\end{figure}

\begin{figure}[H]
\begin{center}
\includegraphics[width=\columnwidth]{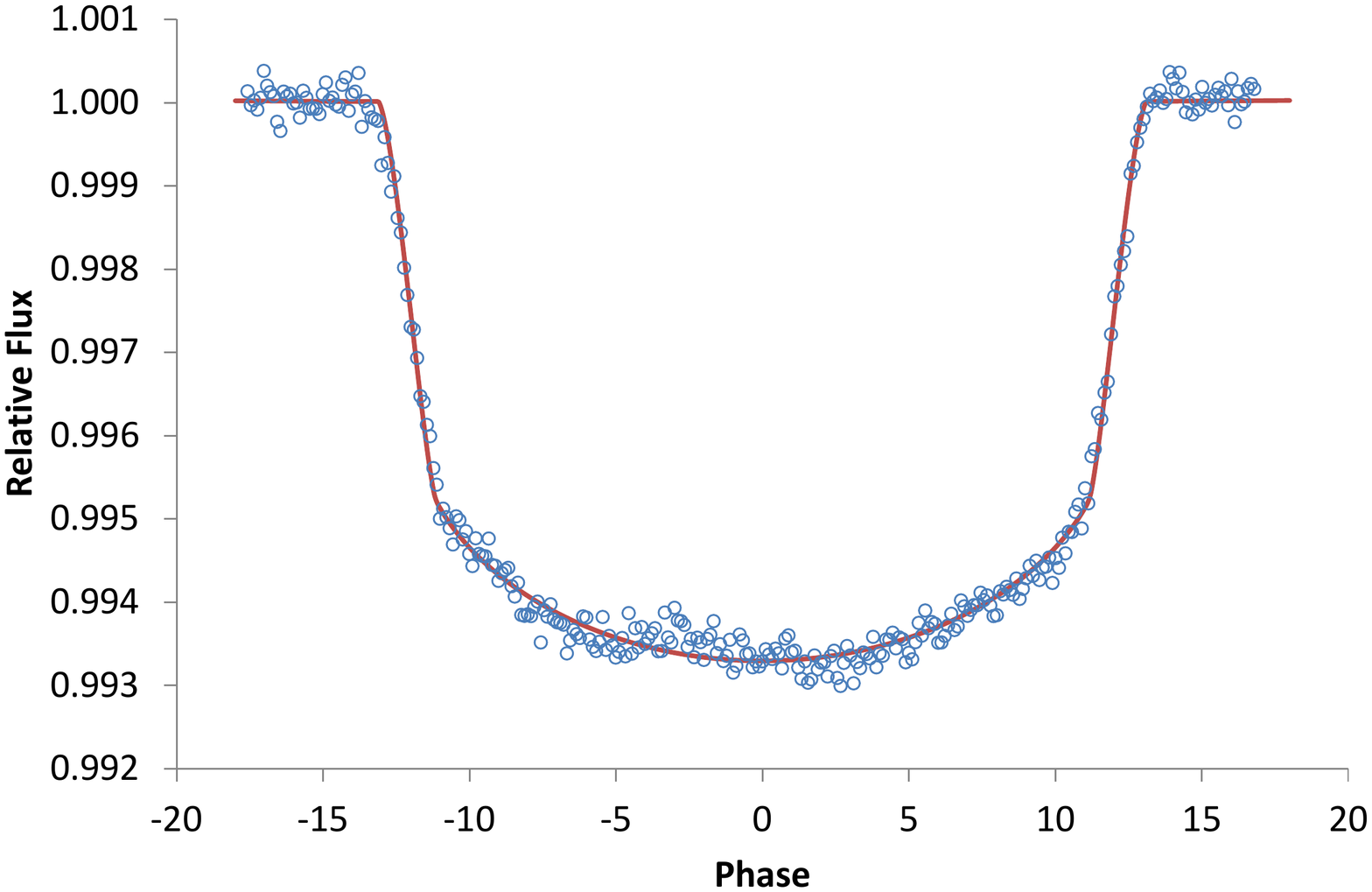} 
\caption{NEA light curve of KOI 2.01, and its {\sc WinKepler} model.}
\label{fig:KOI2}
\end{center}
\end{figure} 

\begin{figure}[H]
\begin{center}
\includegraphics[width=\columnwidth]{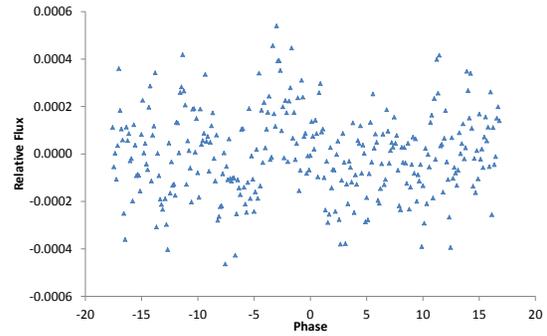} 
\caption{The residuals for KOI 2.01. The starspot effect at phases --5-0 deg is clearly noticed.}
\label{fig:KOI2d}
\end{center}
\end{figure}

\begin{figure}[H]
\begin{center}
\includegraphics[width=\columnwidth]{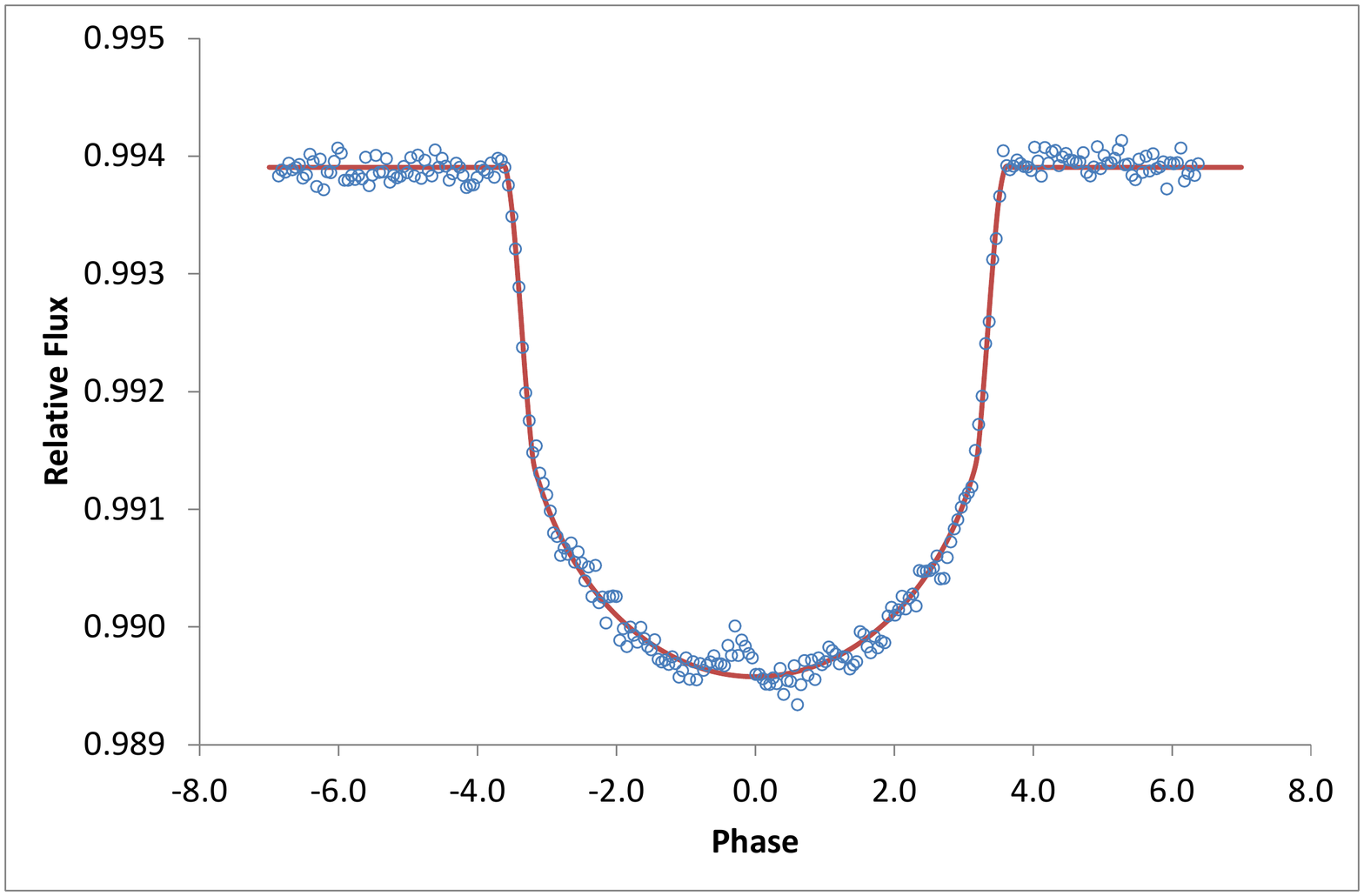} 
\caption{NEA light curve of KOI 3.01, and its {\sc WinKepler} model.}
\label{fig:KOI3}
\end{center}
\end{figure} 

\begin{figure}[H]
\begin{center}
\includegraphics[width=\columnwidth]{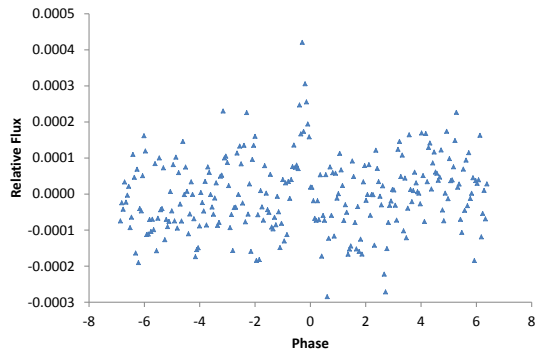} 
\caption{Residuals for KOI 3.01. Again, the starspot just below phase zero
is prominent in the difference curve. }
\label{fig:KOI3d}
\end{center}
\end{figure}

\begin{figure}[H]
\begin{center}
\includegraphics[width=\columnwidth]{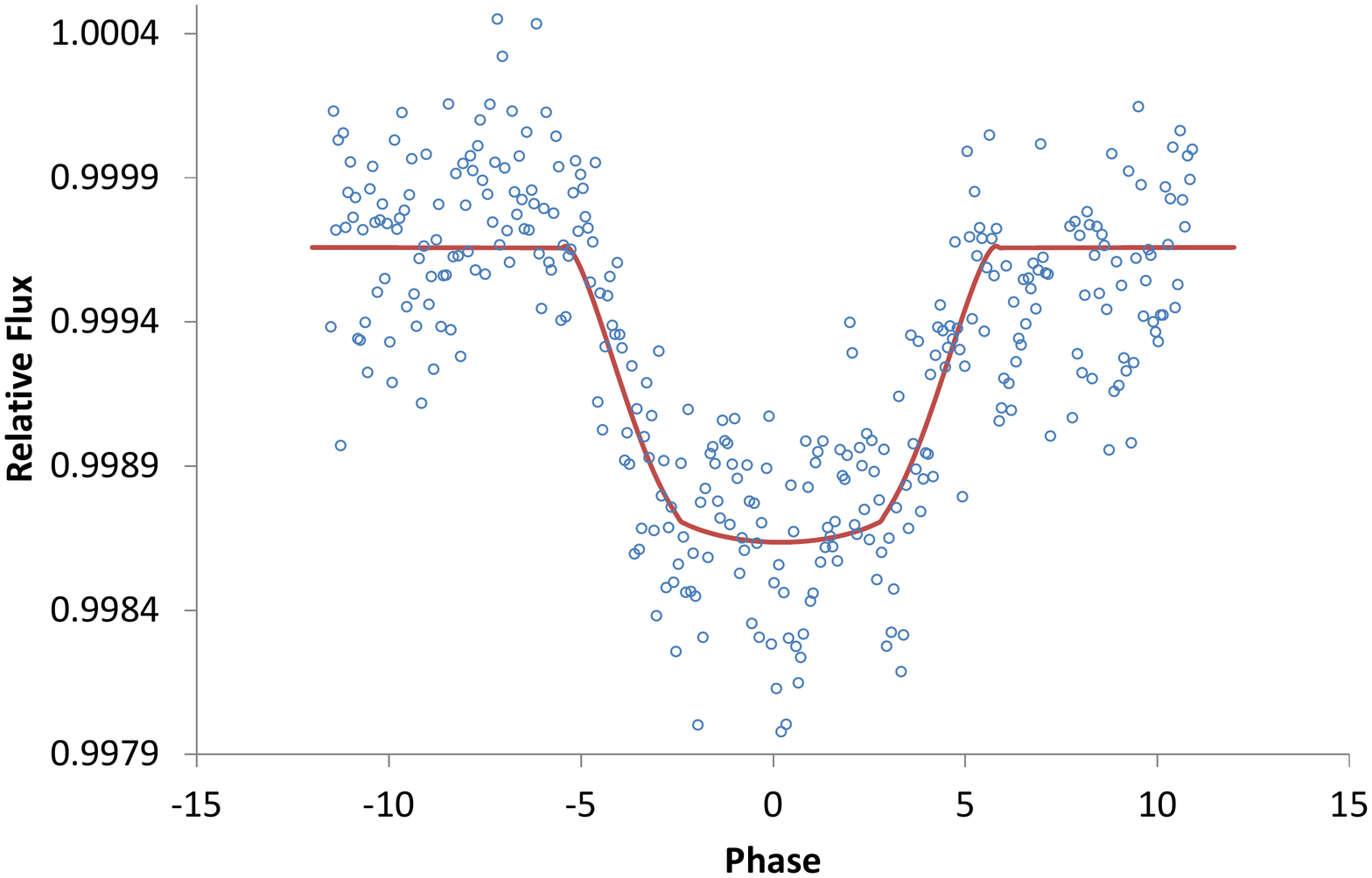} 
\caption{NEA light curve of KOI 4.01, and its {\sc WinKepler} model.}
\label{fig:KOI4}
\end{center}
\end{figure} 

\begin{figure}[H]
\begin{center}
\includegraphics[width=\columnwidth]{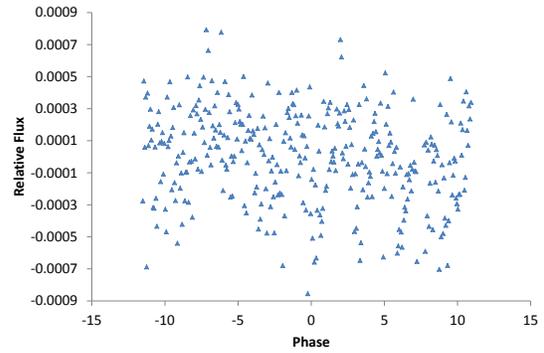} 
\caption{The residuals for KOI 4.01.
This transit may also be affected by a background maculation effect: there appears a slight but steady dimming trend in the difference curve.}
\label{fig:KOI4d}
\end{center}
\end{figure}

\begin{figure}[H]
\begin{center}
\includegraphics[width=\columnwidth]{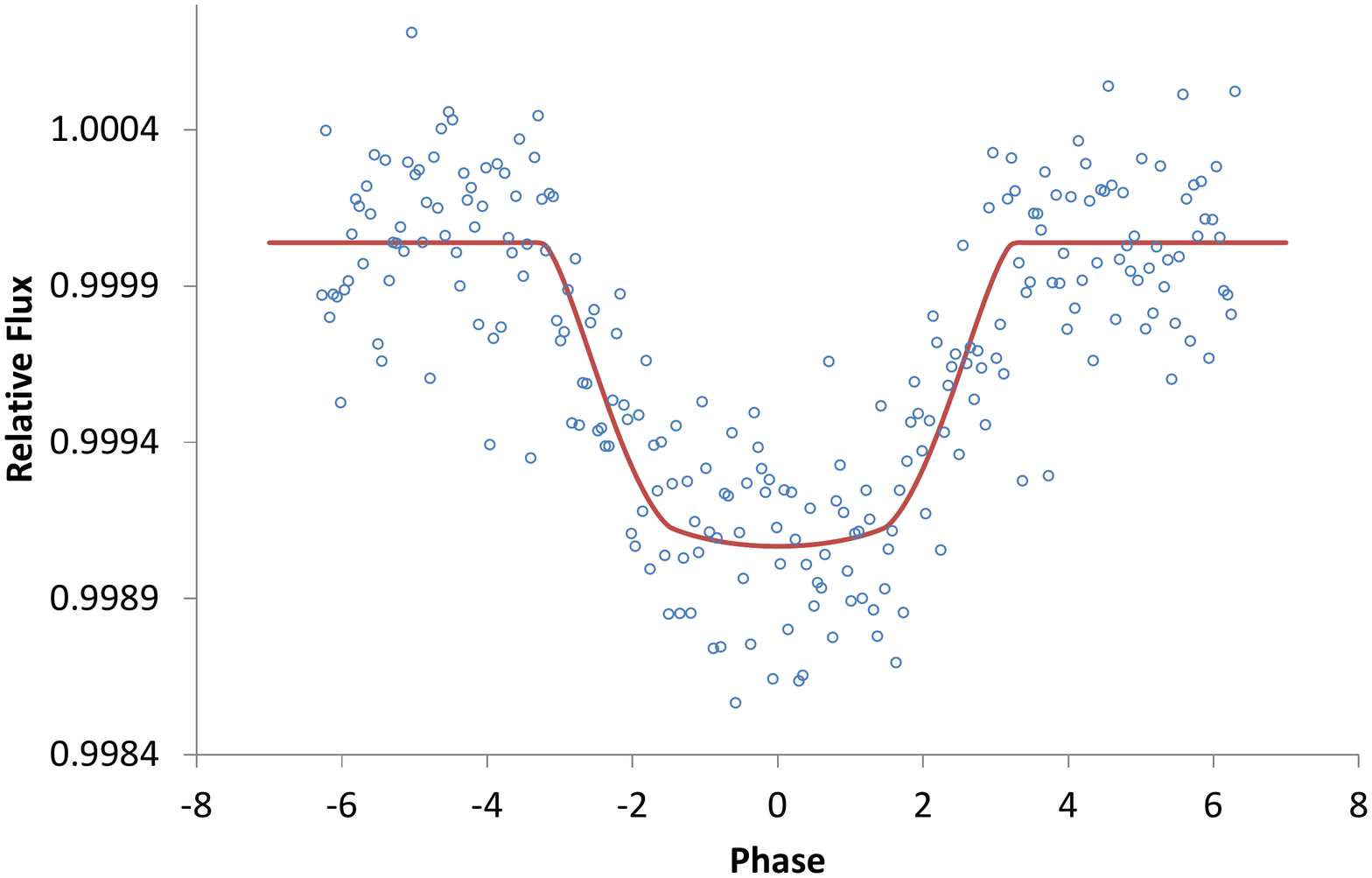} 
\caption{NEA light curve of KOI 5.01, and its {\sc WinKepler} model.}
\label{fig:KOI5}
\end{center}
\end{figure} 

\begin{figure}[H]
\begin{center}
\includegraphics[width=\columnwidth]{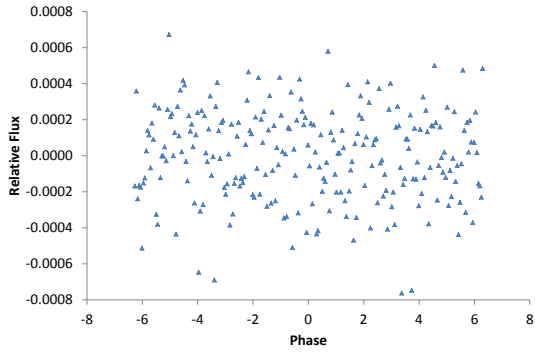} 
\caption{Residuals for KOI 5.01}
\label{fig:KOI5d}
\end{center}
\end{figure}

\begin{figure}[H]
\begin{center}
\includegraphics[width=\columnwidth]{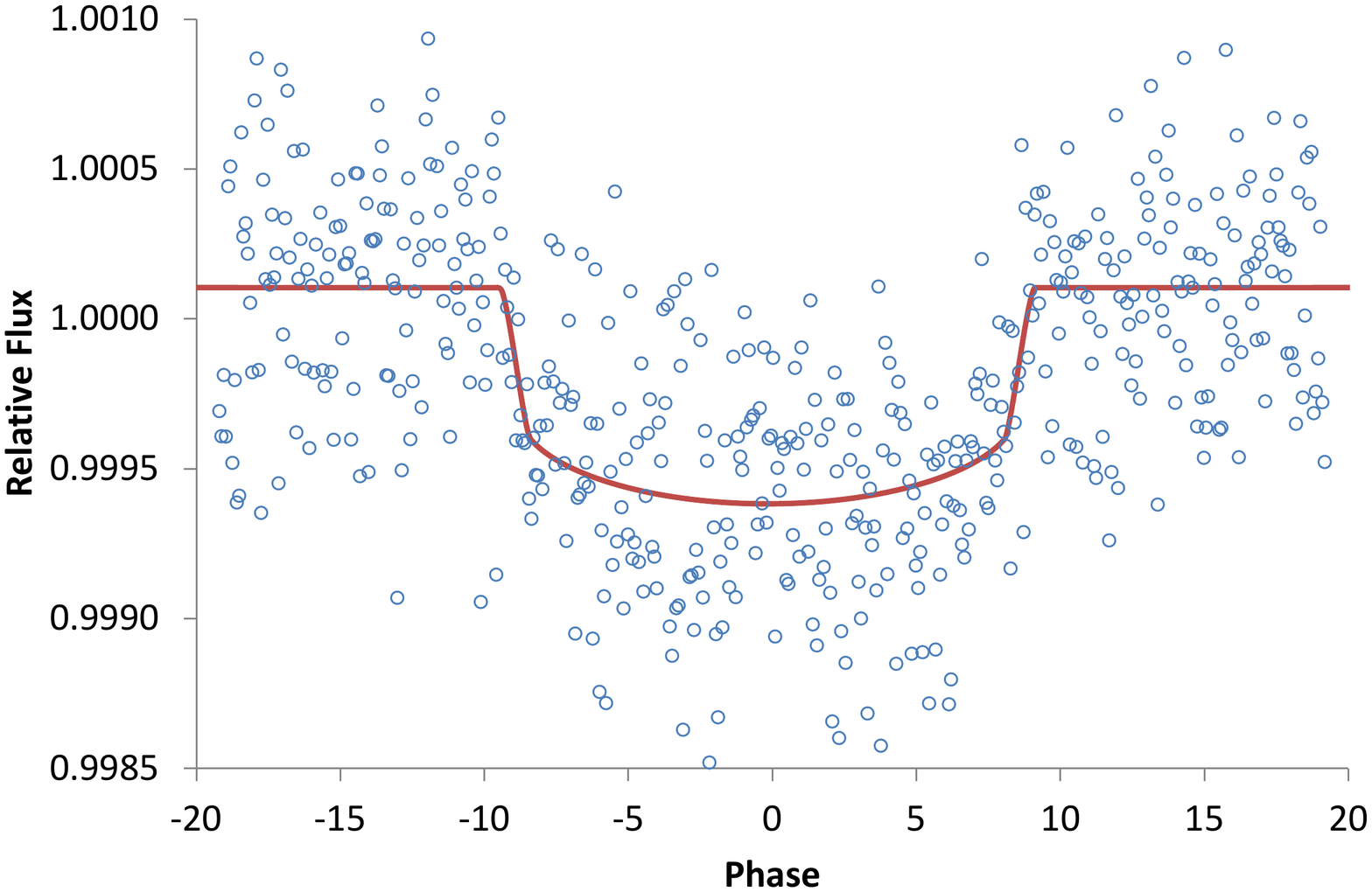} 
\caption{NEA light curve of KOI 7.01, and its {\sc WinKepler} model.}
\label{fig:KOI7}
\end{center}
\end{figure} 

\begin{figure}[H]
\begin{center}
\includegraphics[width=\columnwidth]{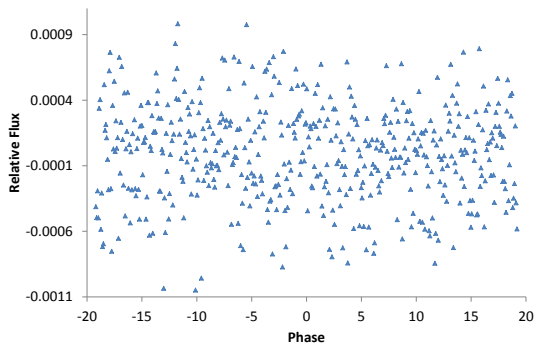} 
\caption{The residuals for KOI 7.01. Note the relatively large, though uniform, scatter.}
\label{fig:KOI7d}
\end{center}
\end{figure}

\begin{figure}[H]
\begin{center}
\includegraphics[width=\columnwidth]{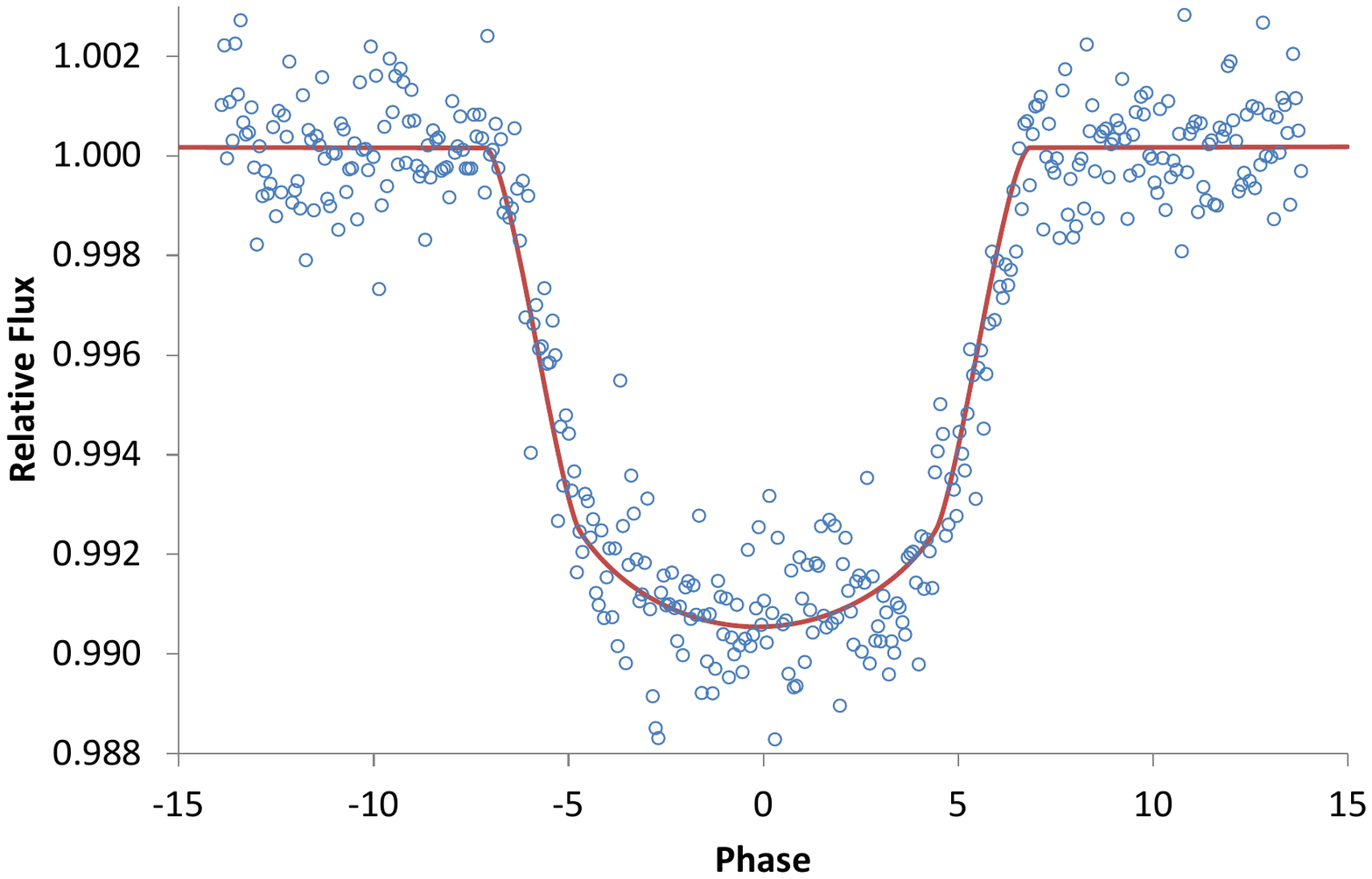} 
\caption{NEA light curve of KOI 10.01, and its {\sc WinKepler} model.}
\label{fig:KOI10}
\end{center}
\end{figure} 

\begin{figure}[H]
\begin{center}
\includegraphics[width=\columnwidth]{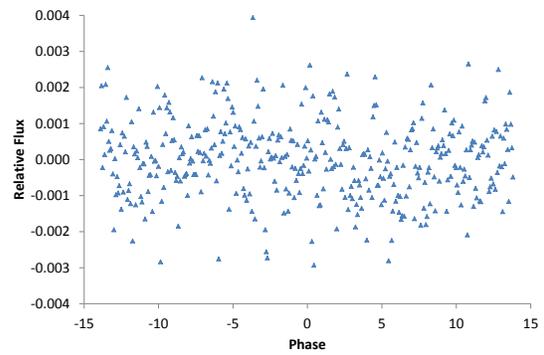} 
\caption{The residuals for KOI 10.01. Note the relatively large scatter.}
\label{fig:KOI10d}
\end{center}
\end{figure}

\begin{figure}[H]
\begin{center}
\includegraphics[width=\columnwidth]{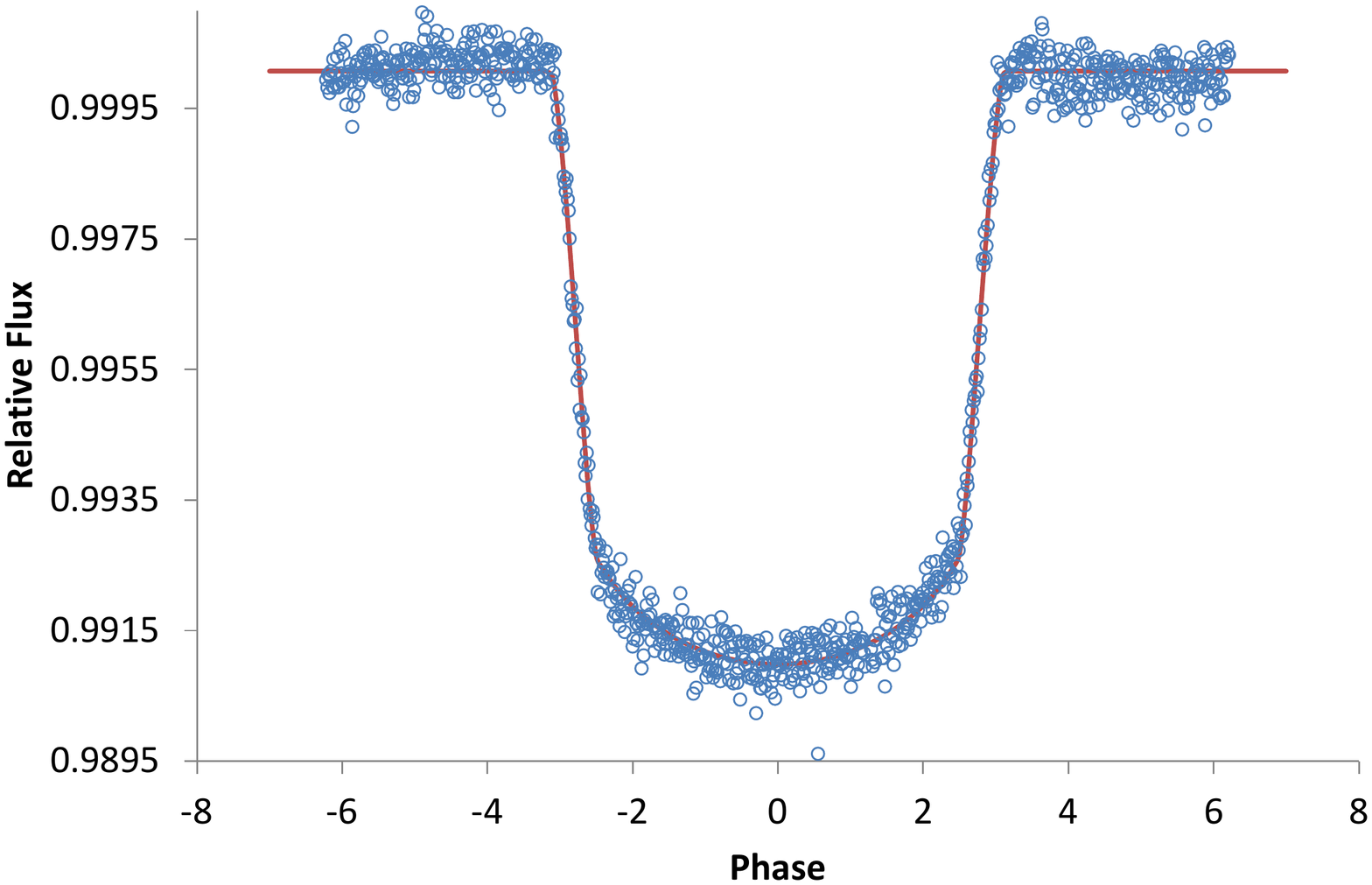} 
\caption{NEA light curve of KOI 12.01, and its {\sc WinKepler} model.}
\label{fig:KOI12}
\end{center}
\end{figure} 

\begin{figure}[H]
\begin{center}
\includegraphics[width=\columnwidth]{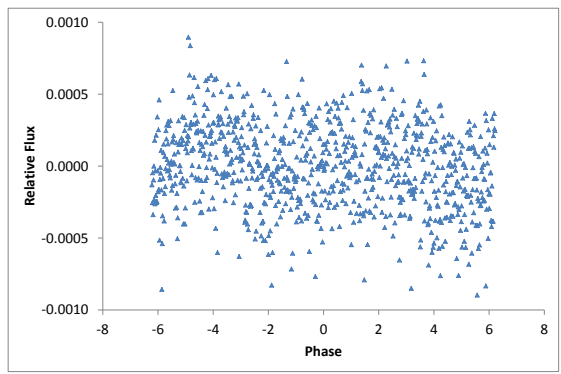} 
\caption{The residuals for KOI 12.01, with optimized limb-darkening coefficient}
\label{fig:KOI12d}
\end{center}
\end{figure}
 
\begin{figure}[H]
\begin{center}
\includegraphics[width=\columnwidth]{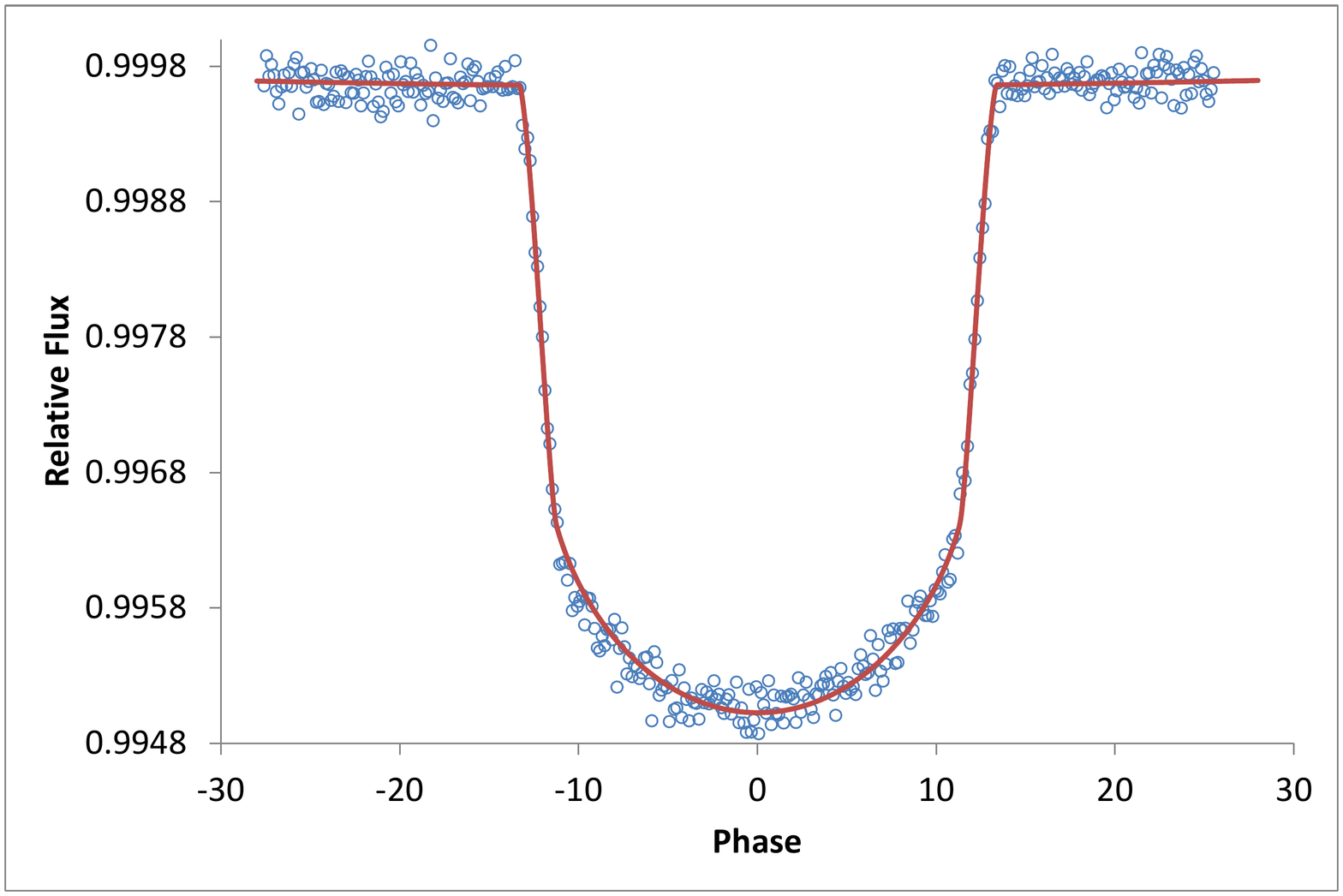} 
\caption{NEA light curve of KOI 13.01, and its {\sc WinKepler} model.}
\label{fig:KOI13}
\end{center}
\end{figure} 

\begin{figure}[H]
\begin{center}
\includegraphics[width=\columnwidth]{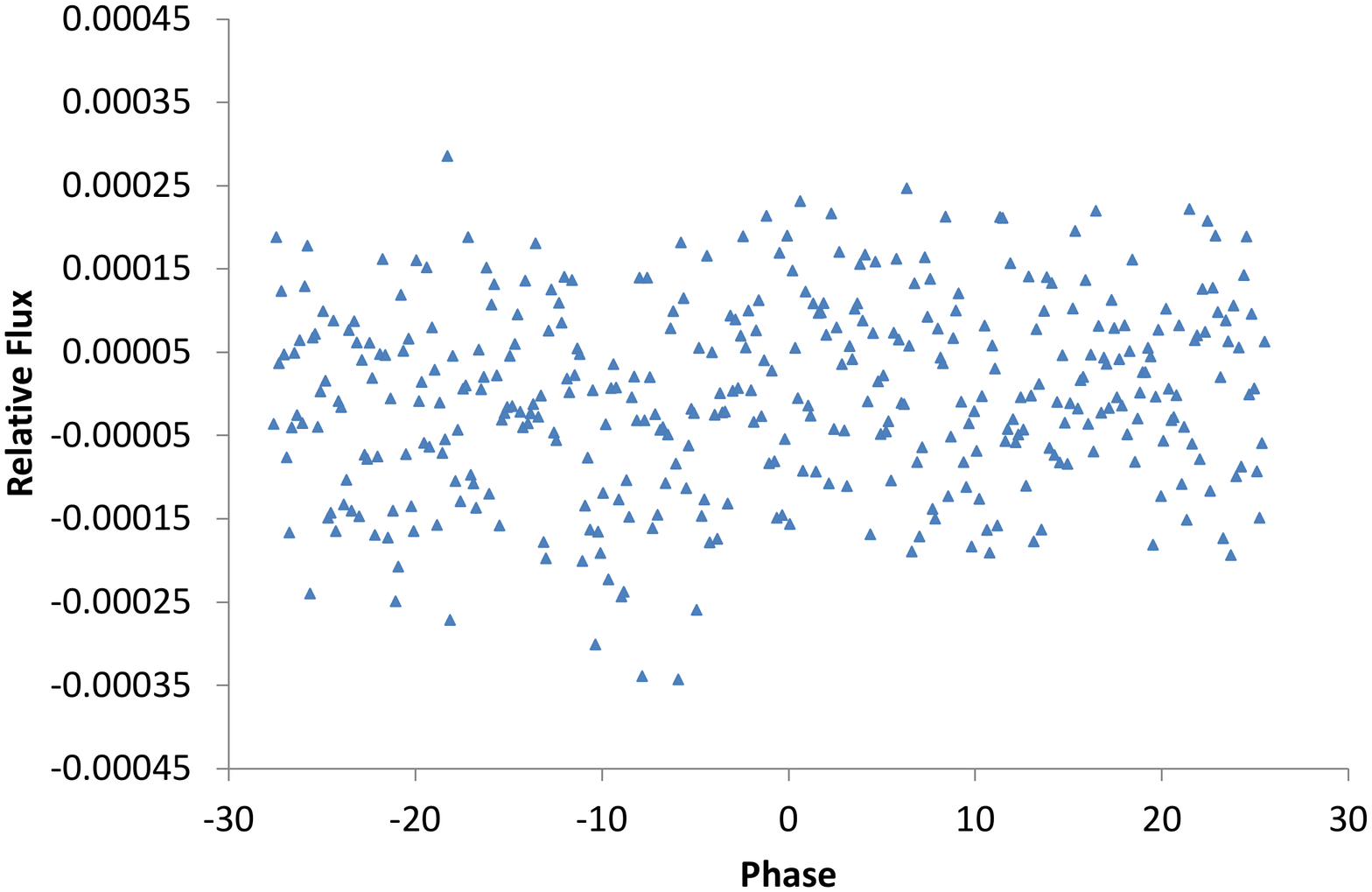} 
\caption{The residuals for KOI 13.01}
\label{fig:KOI13d}
\end{center}
\end{figure}

\begin{figure}[H]
\begin{center}
\includegraphics[width=\columnwidth]{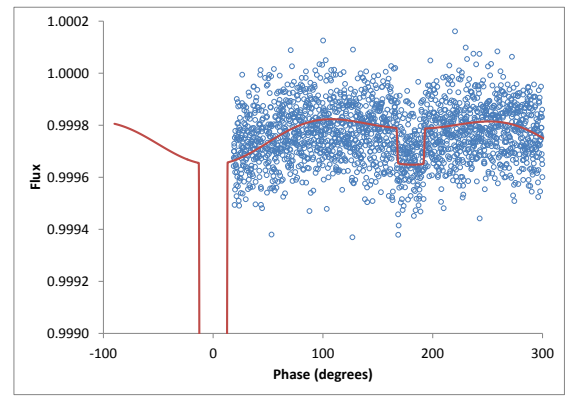} 
\caption{The out-of-transit region of the light curve of KOI 13}
\label{fig:KOI13ecl}
\end{center}
\end{figure}

\begin{figure}[H]
\begin{center}
\includegraphics[width=\columnwidth]{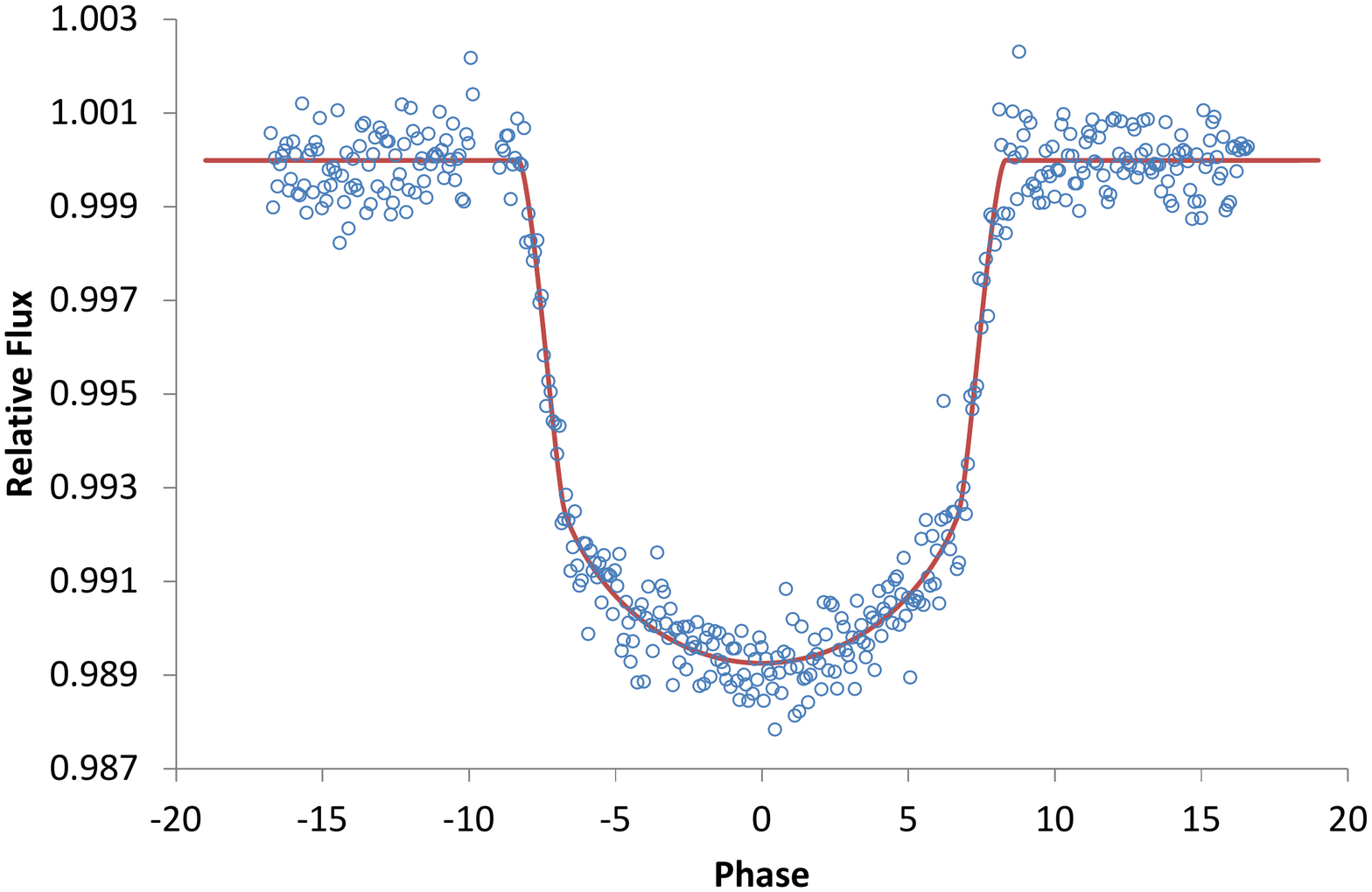} 
\caption{NEA light curve of KOI 17.01, and its {\sc WinKepler} model.}
\label{fig:KOI17}
\end{center}
\end{figure} 

\begin{figure}[H]
\begin{center}
\includegraphics[width=\columnwidth]{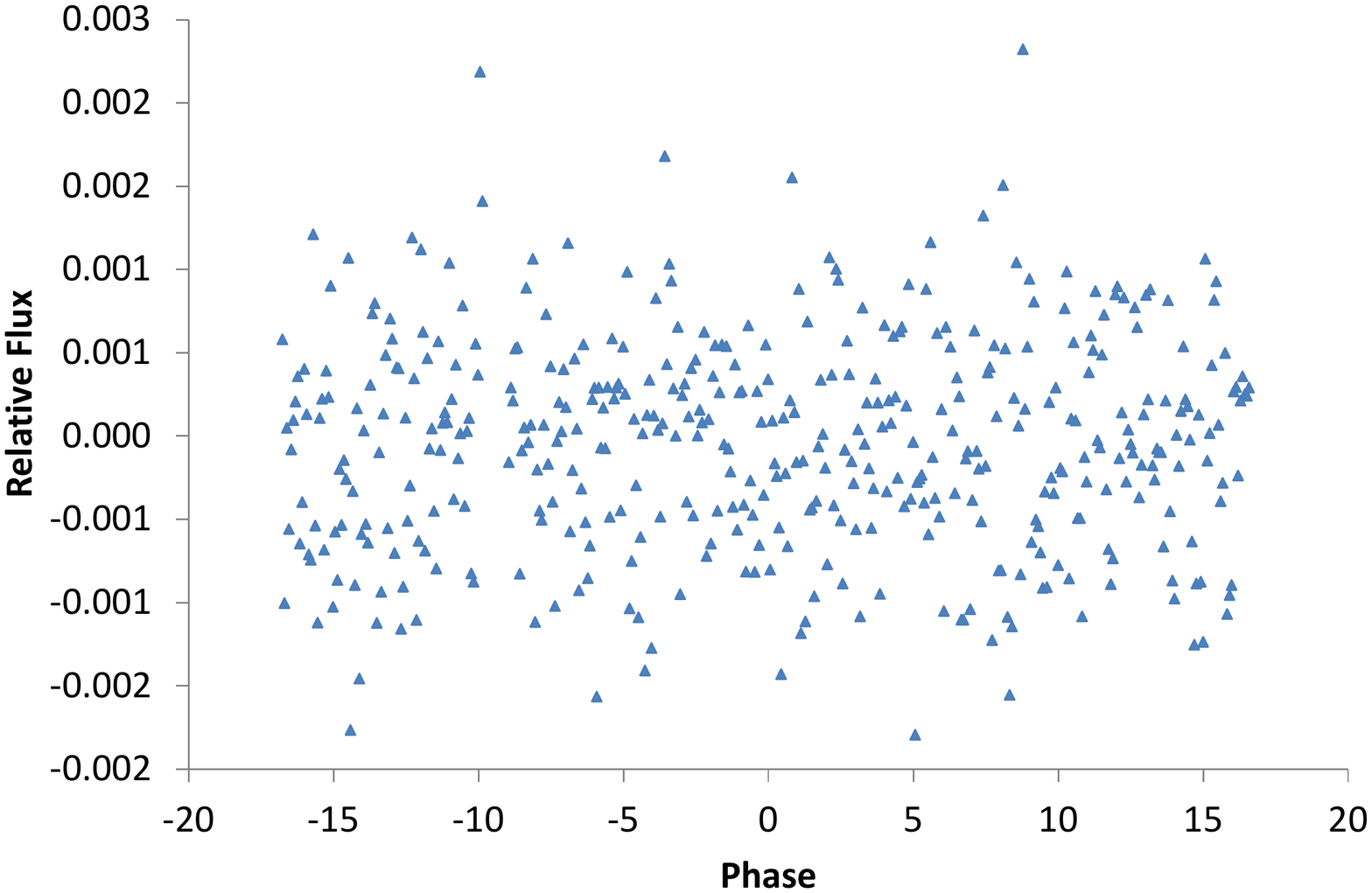} 
\caption{The residuals for KOI 17.01}
\label{fig:KOI17d}
\end{center}
\end{figure}

\begin{figure}[H]
\begin{center}
\includegraphics[width=\columnwidth]{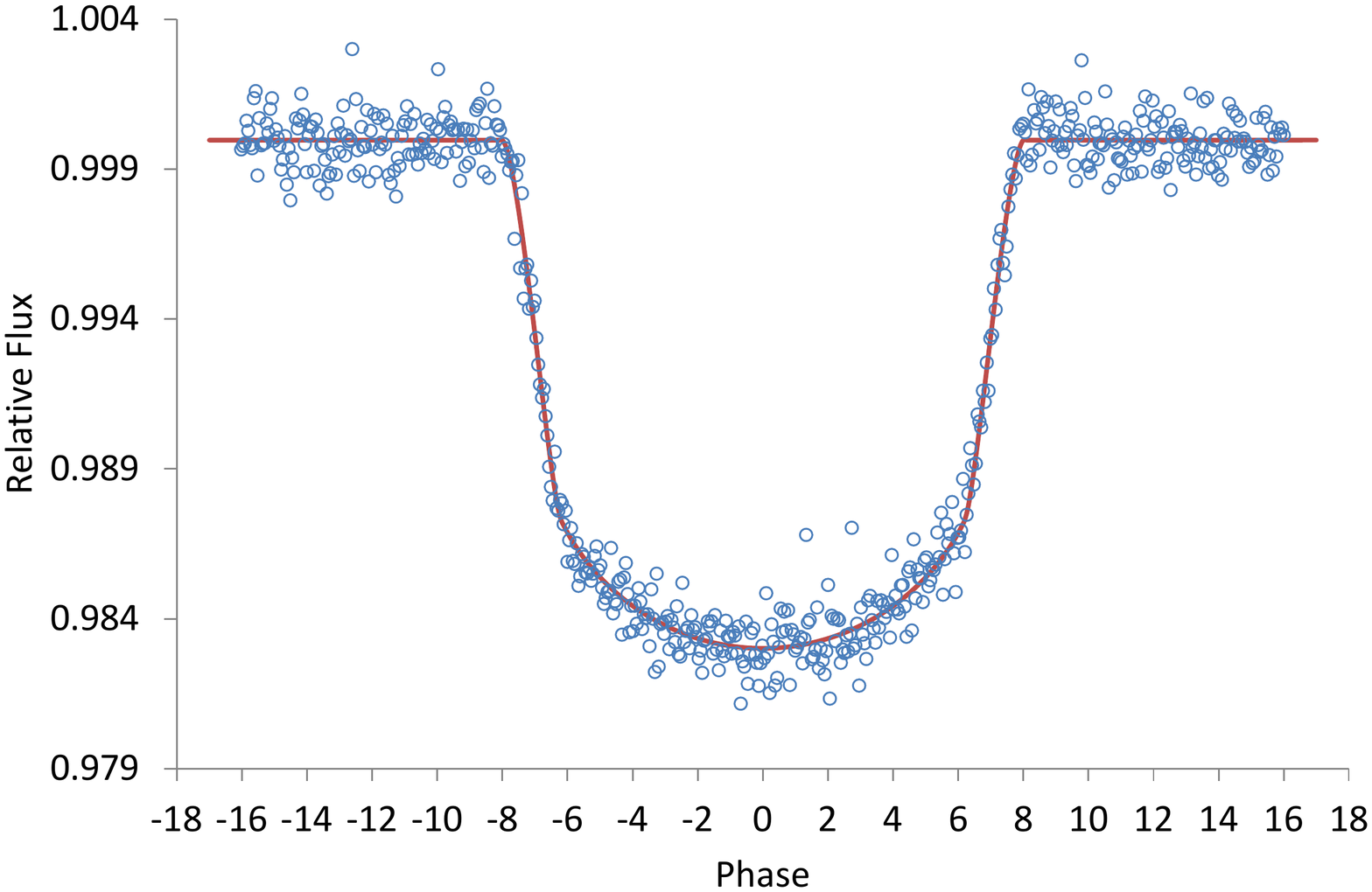} 
\caption{NEA light curve of KOI 20.01, and its {\sc WinKepler} model.}
\label{fig:KOI20.01}
\end{center}
\end{figure} 

\begin{figure}[H]
\begin{center}
\includegraphics[width=\columnwidth]{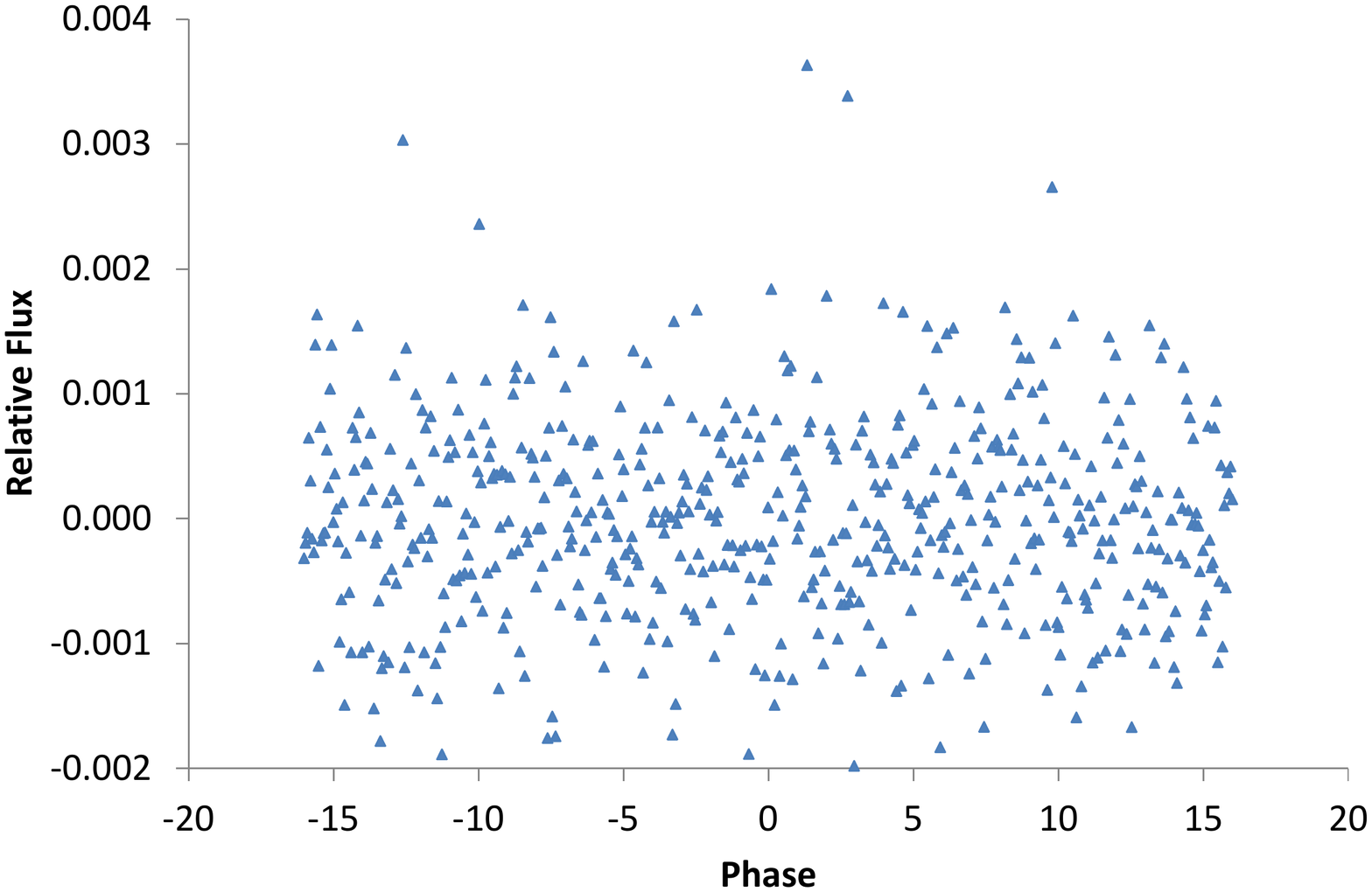} 
\caption{The residuals for KOI 20.01}
\label{fig:KOI20.01d}
\end{center}
\end{figure}

\begin{figure}[H]
\begin{center}
\includegraphics[width=\columnwidth]{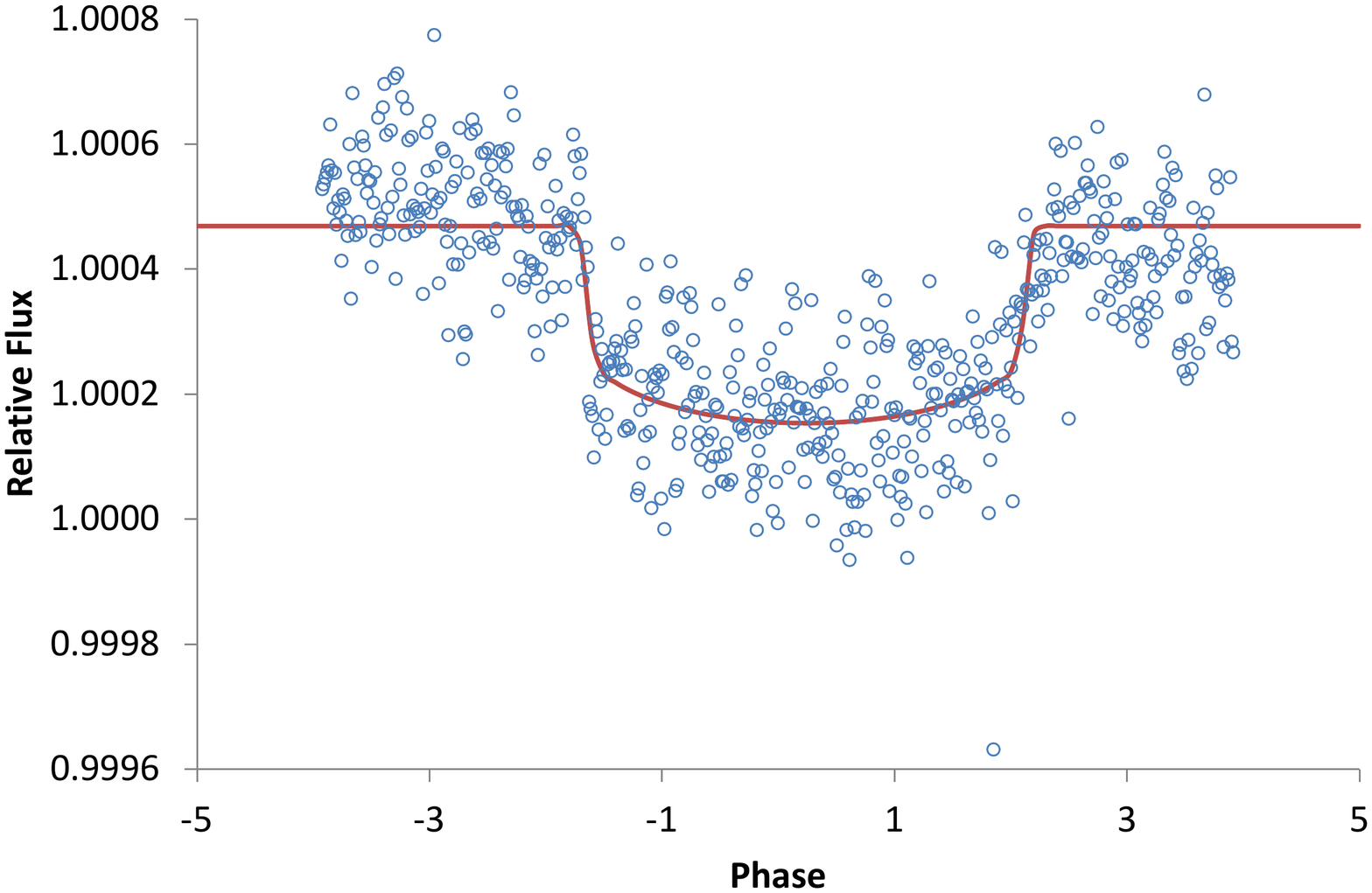} 
\caption{NEA light curve of KOI 42.01, and its {\sc WinKepler} model.}
\label{fig:KOI42}
\end{center}
\end{figure} 

\begin{figure}[H]
\begin{center}
\includegraphics[width=\columnwidth]{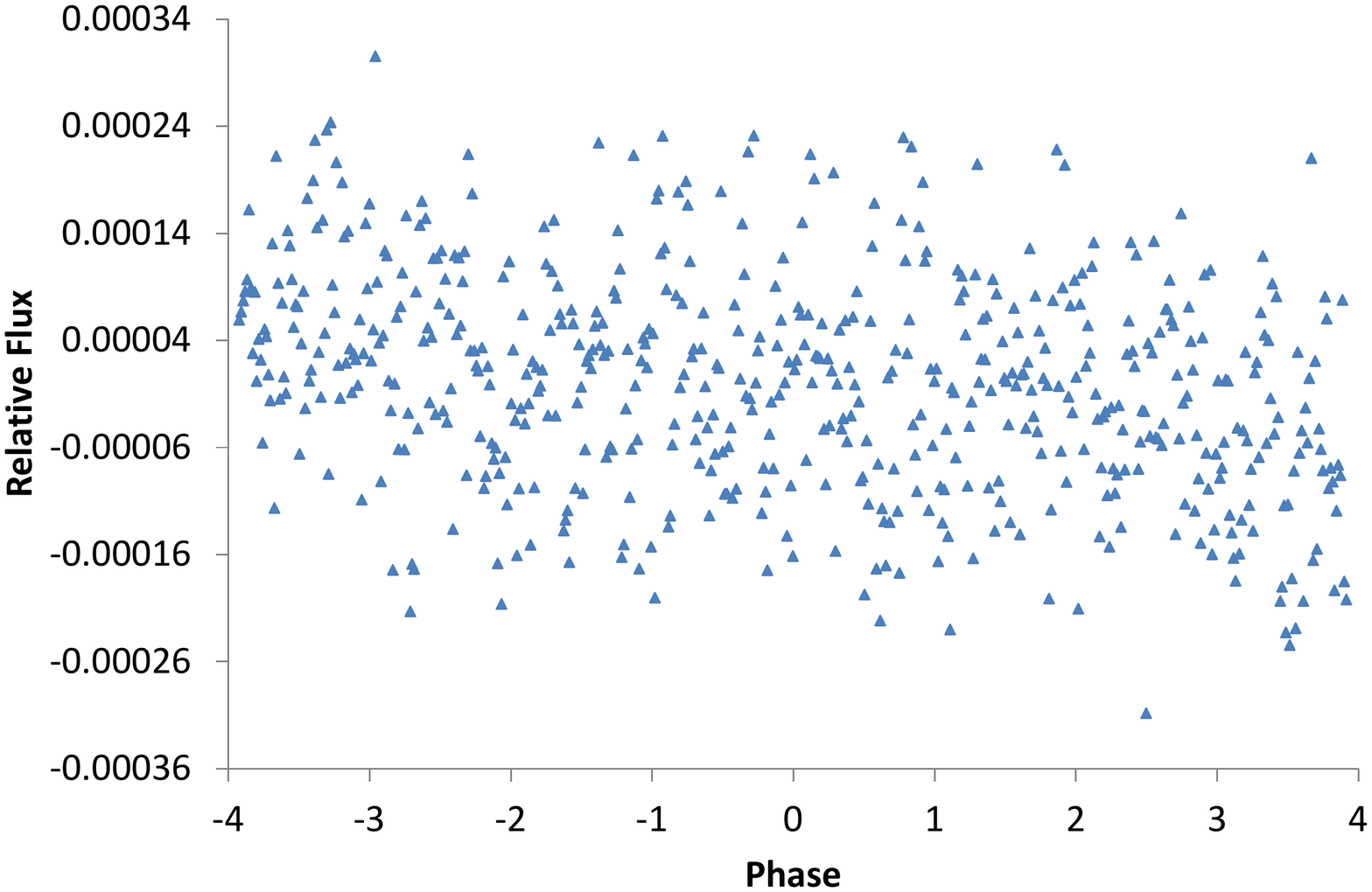} 
\caption{The residuals for KOI 42.01}
\label{fig:KOI42d}
\end{center}
\end{figure}

\begin{figure}[H]
\begin{center}
\includegraphics[width=\columnwidth]{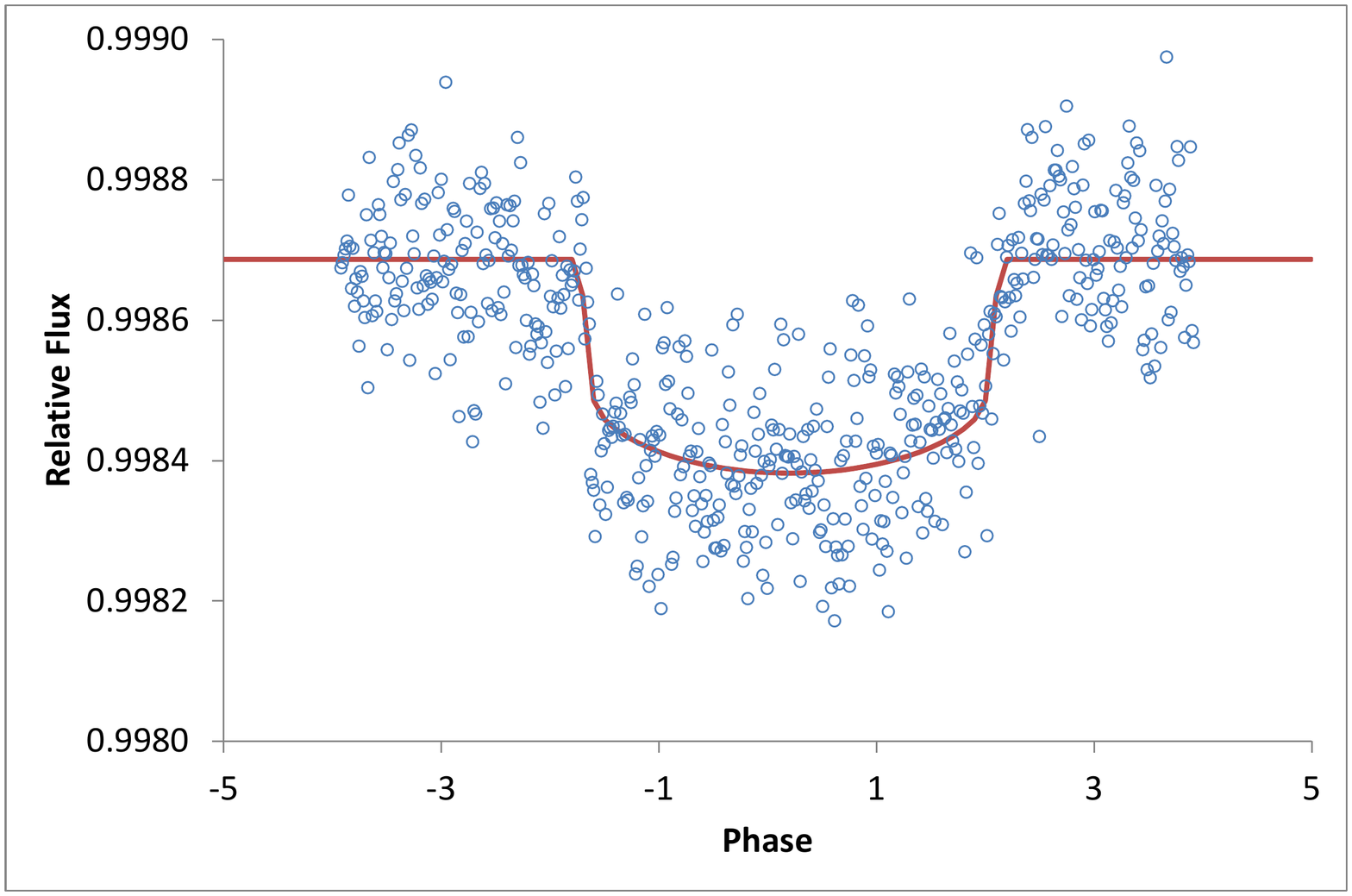} 
\caption{NEA light curve of KOI 42.01(d), and its {\sc WinKepler} model.}
\label{fig:KOI42(d)}
\end{center}
\end{figure} 

\begin{figure}[H]
\begin{center}
\includegraphics[width=\columnwidth]{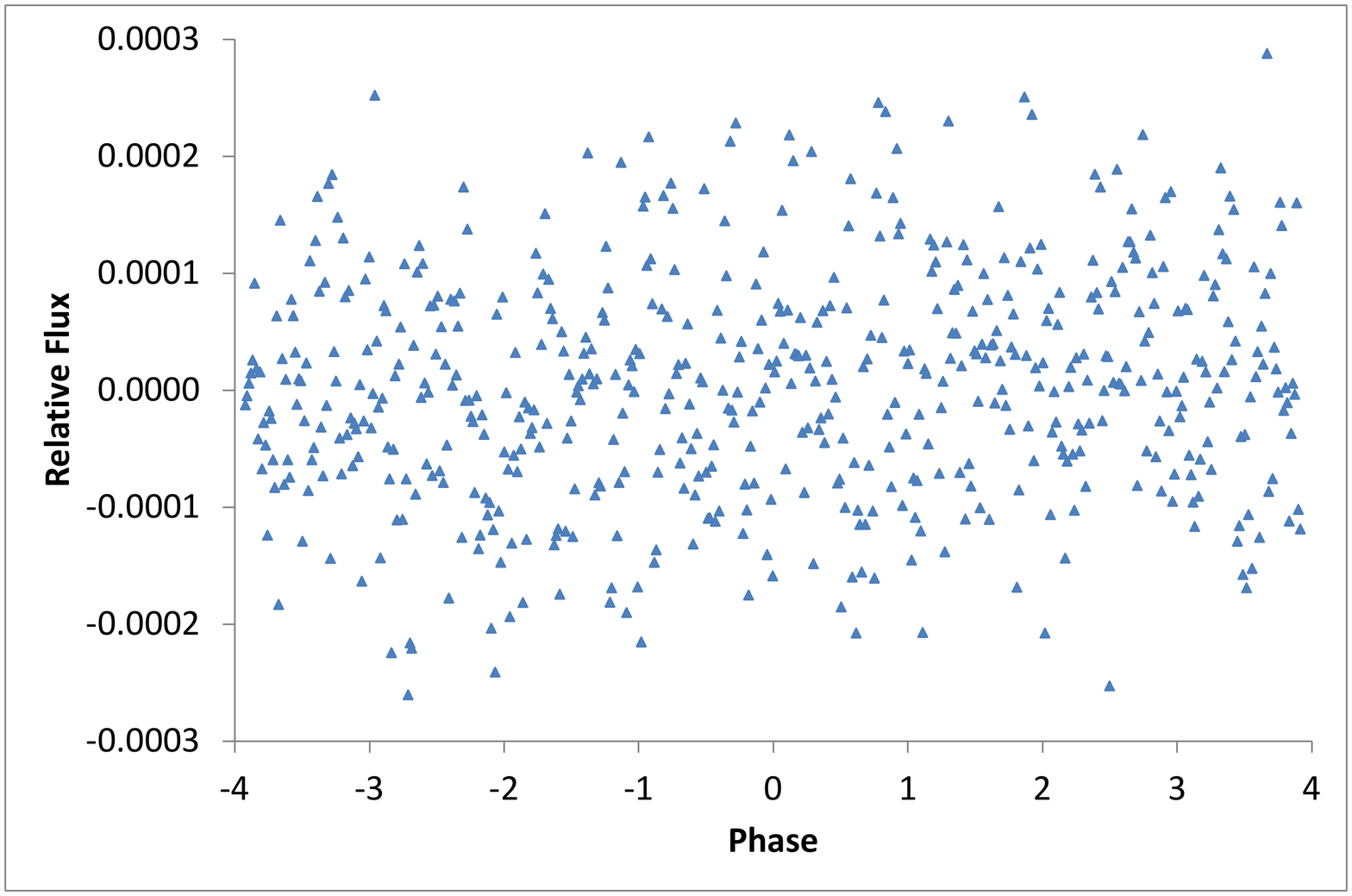} 
\caption{The residuals for KOI 42.01(d)}
\label{fig:KOI42(d)d}
\end{center}
\end{figure}

\begin{figure}[H]
\begin{center}
\includegraphics[width=\columnwidth]{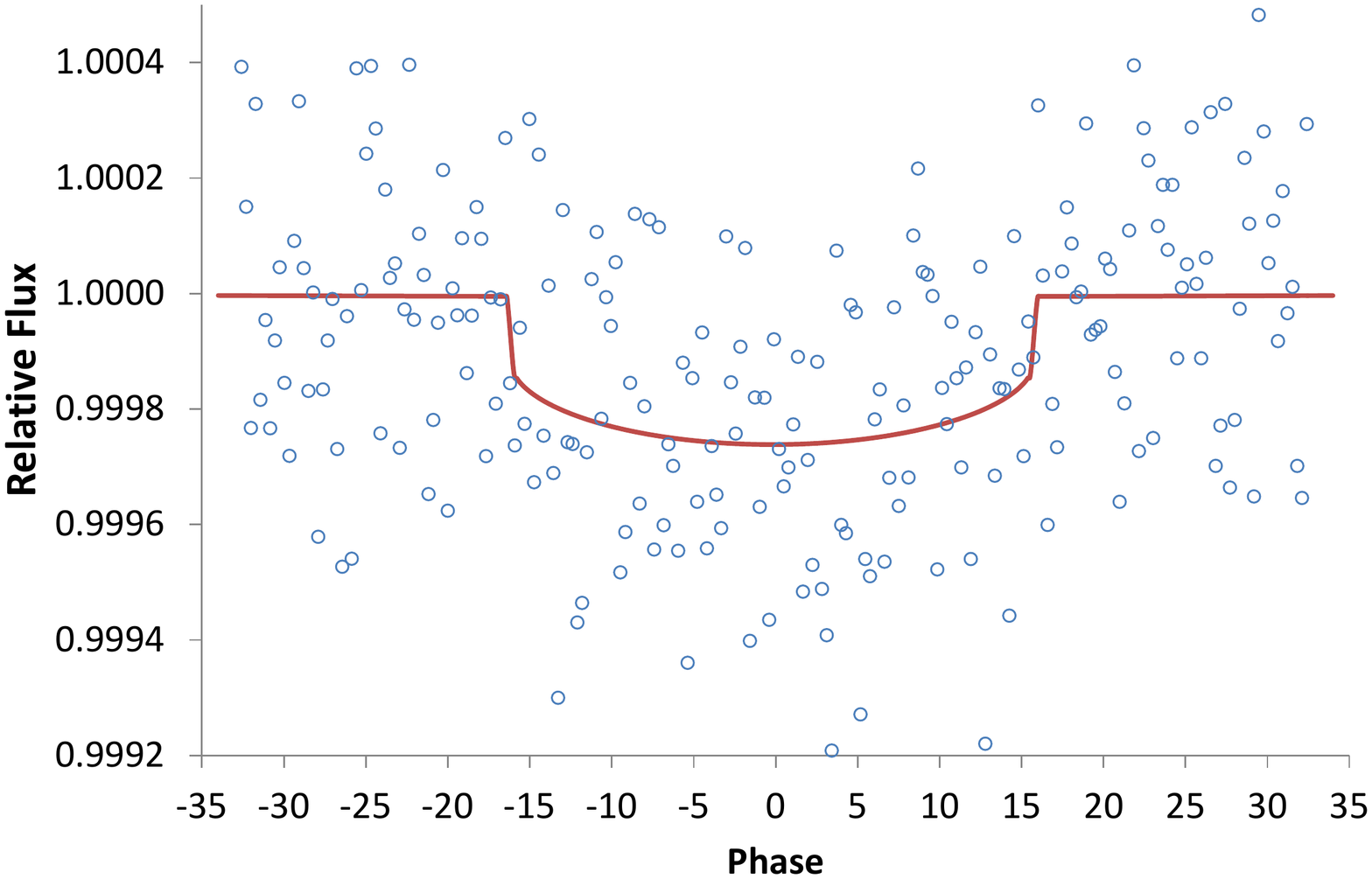} 
\caption{NEA light curve of KOI 72.01, and its {\sc WinKepler} model.}
\label{fig:KOI72}
\end{center}
\end{figure} 

\begin{figure}[H]
\begin{center}
\includegraphics[width=\columnwidth]{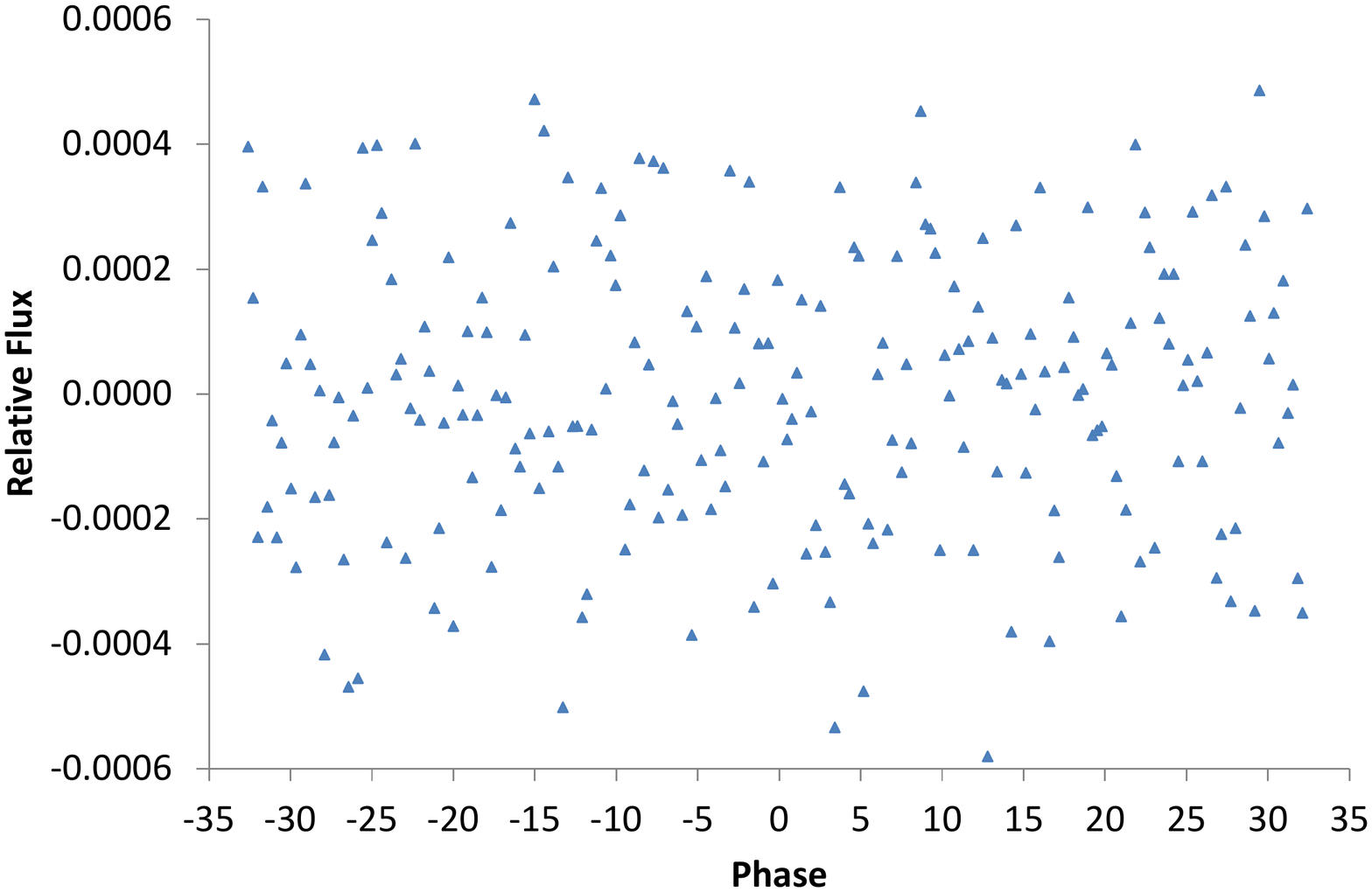} 
\caption{The residuals for KOI 72.01}
\label{fig:KOI72d}
\end{center}
\end{figure}

\begin{figure}[H]
\begin{center}
\includegraphics[width=\columnwidth]{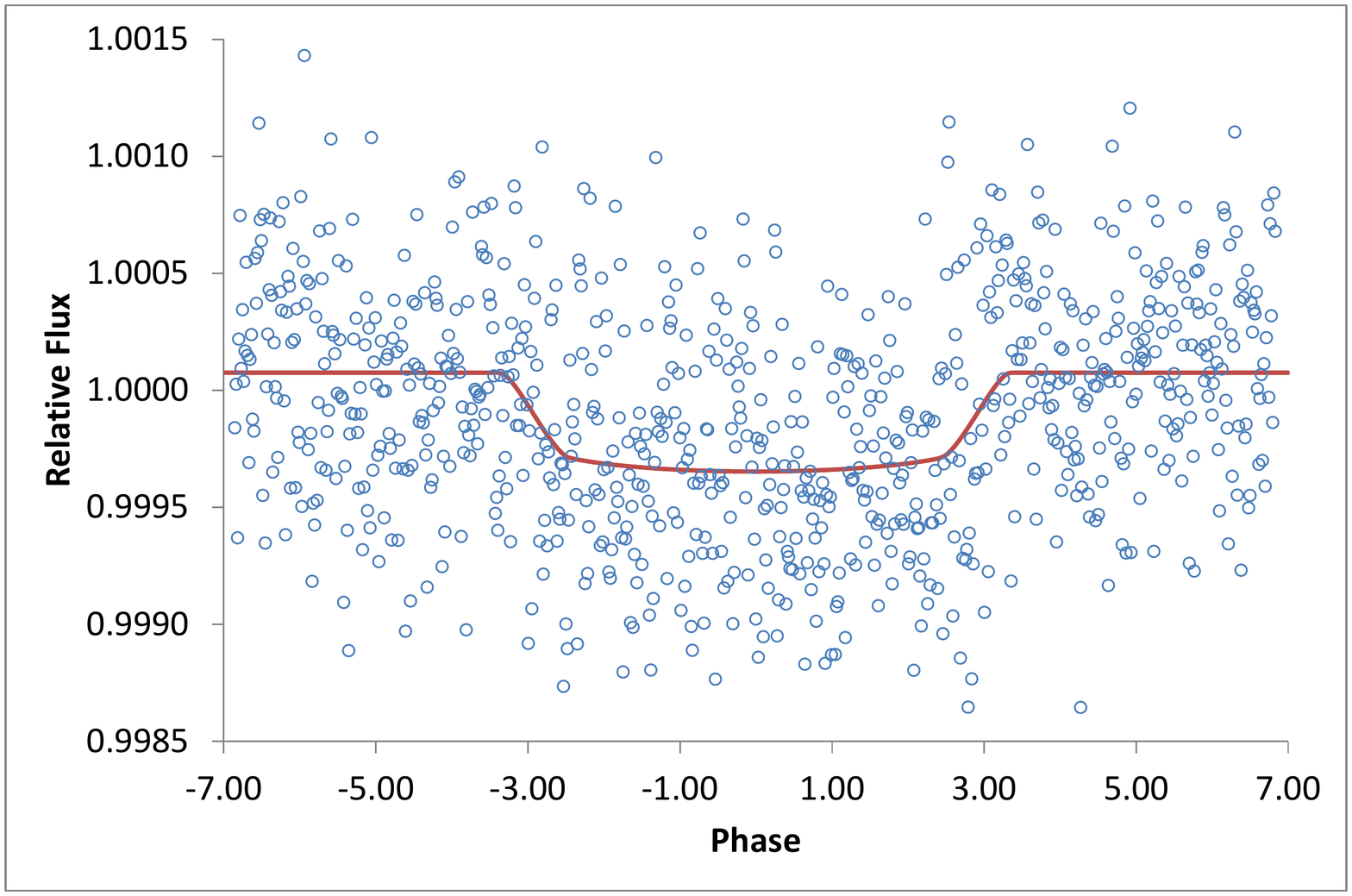} 
\caption{NEA light curve of KOI 117.01, and its {\sc WinKepler} model.}
\label{fig:KOI117}
\end{center}
\end{figure} 

\begin{figure}[H]
\begin{center}
\includegraphics[width=\columnwidth]{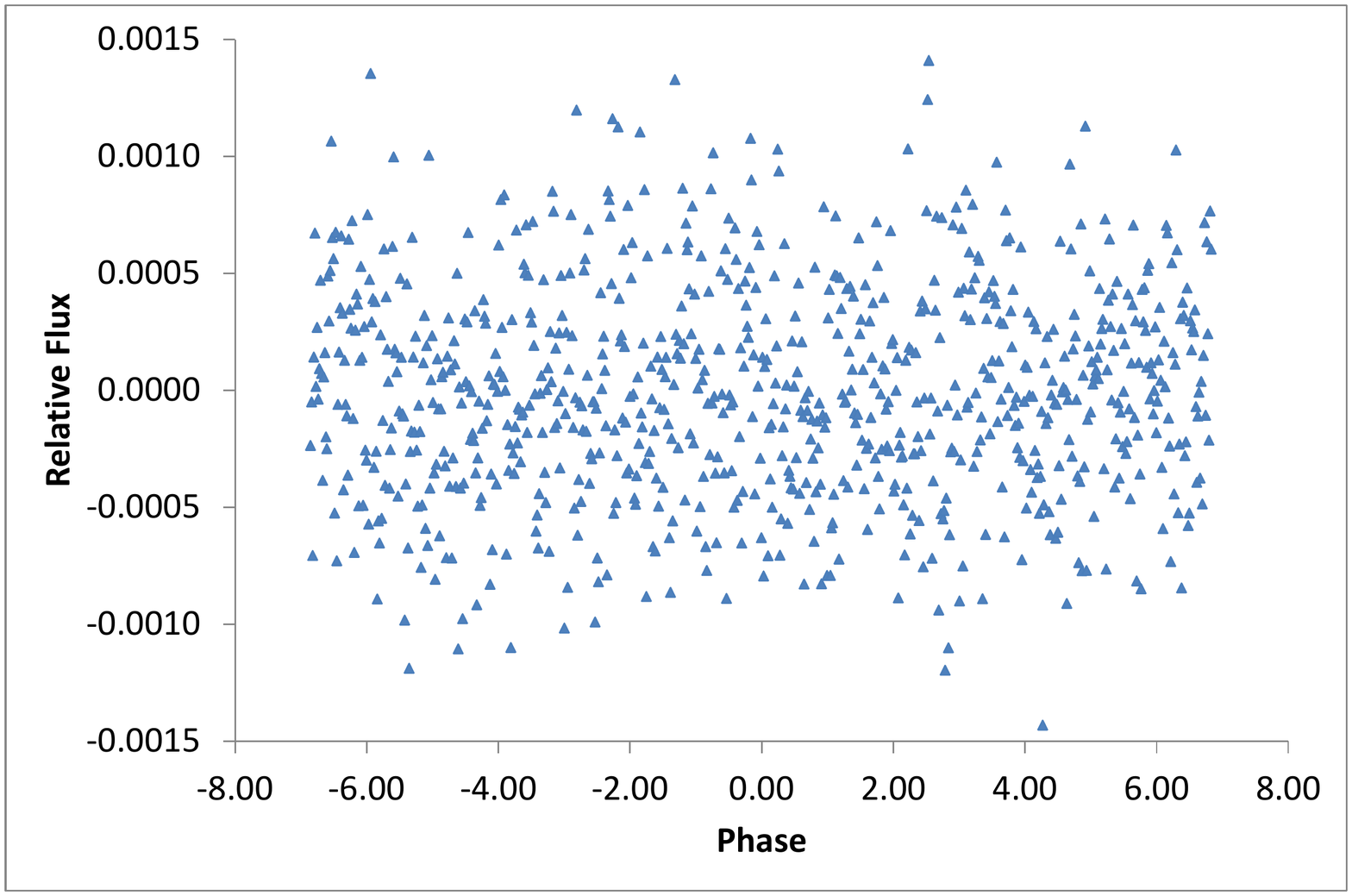} 
\caption{The residuals for KOI 117.01}
\label{fig:KOI117d}
\end{center}
\end{figure}

\begin{figure}[H]
\begin{center}
\includegraphics[width=\columnwidth]{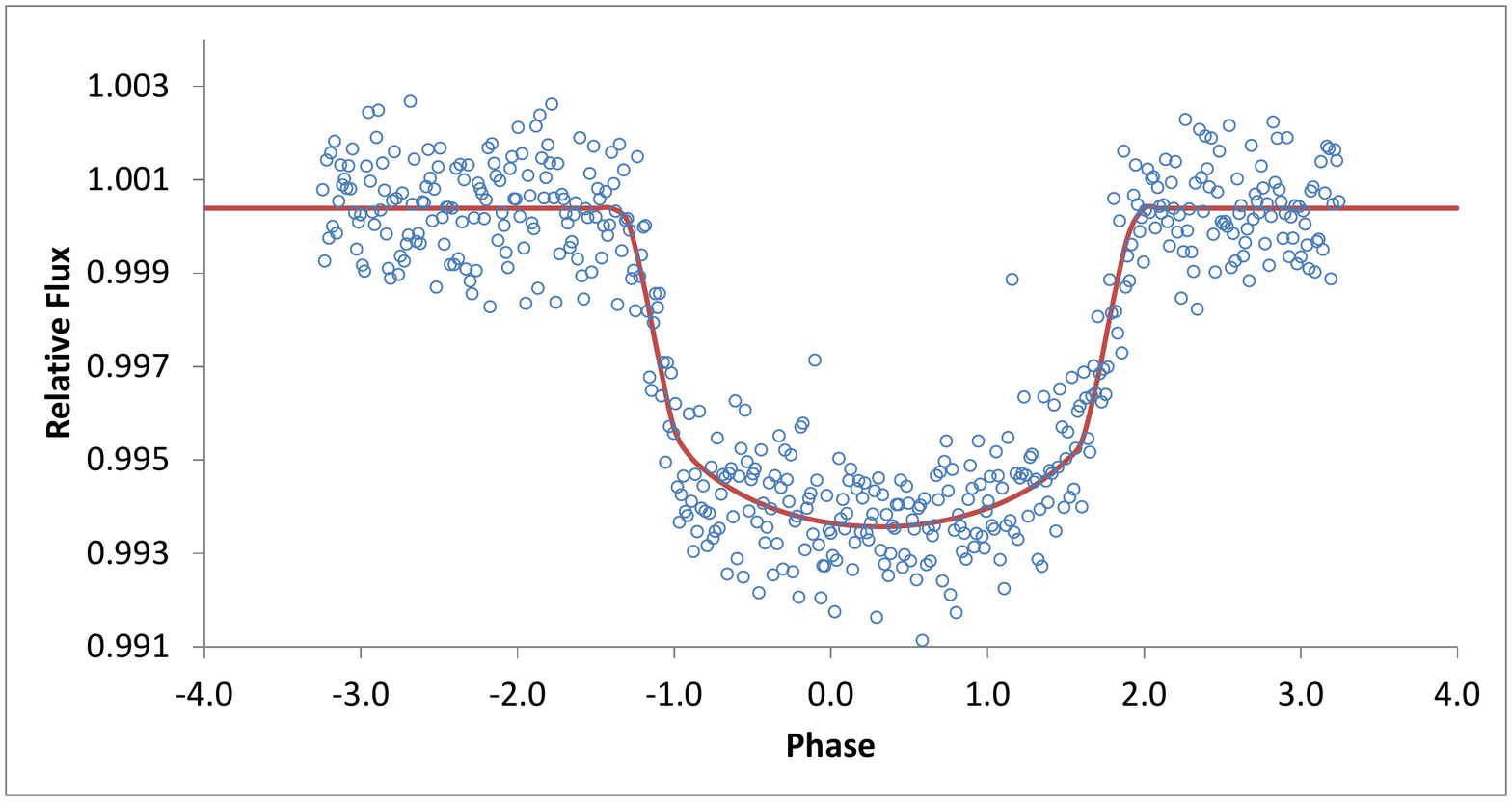} 
\caption{NEA light curve of KOI 377.01, and its {\sc WinKepler} model.}
\label{fig:KOI377}
\end{center}
\end{figure} 

\begin{figure}[H]
\begin{center}
\includegraphics[width=\columnwidth]{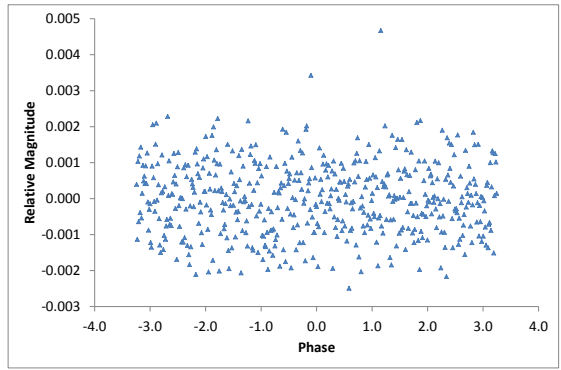} 
\caption{Curve-fitting residuals for KOI 377.01}
\label{fig:KOI377d}
\end{center}
\end{figure}

\begin{figure}[H]
\begin{center}
\includegraphics[width=\columnwidth]{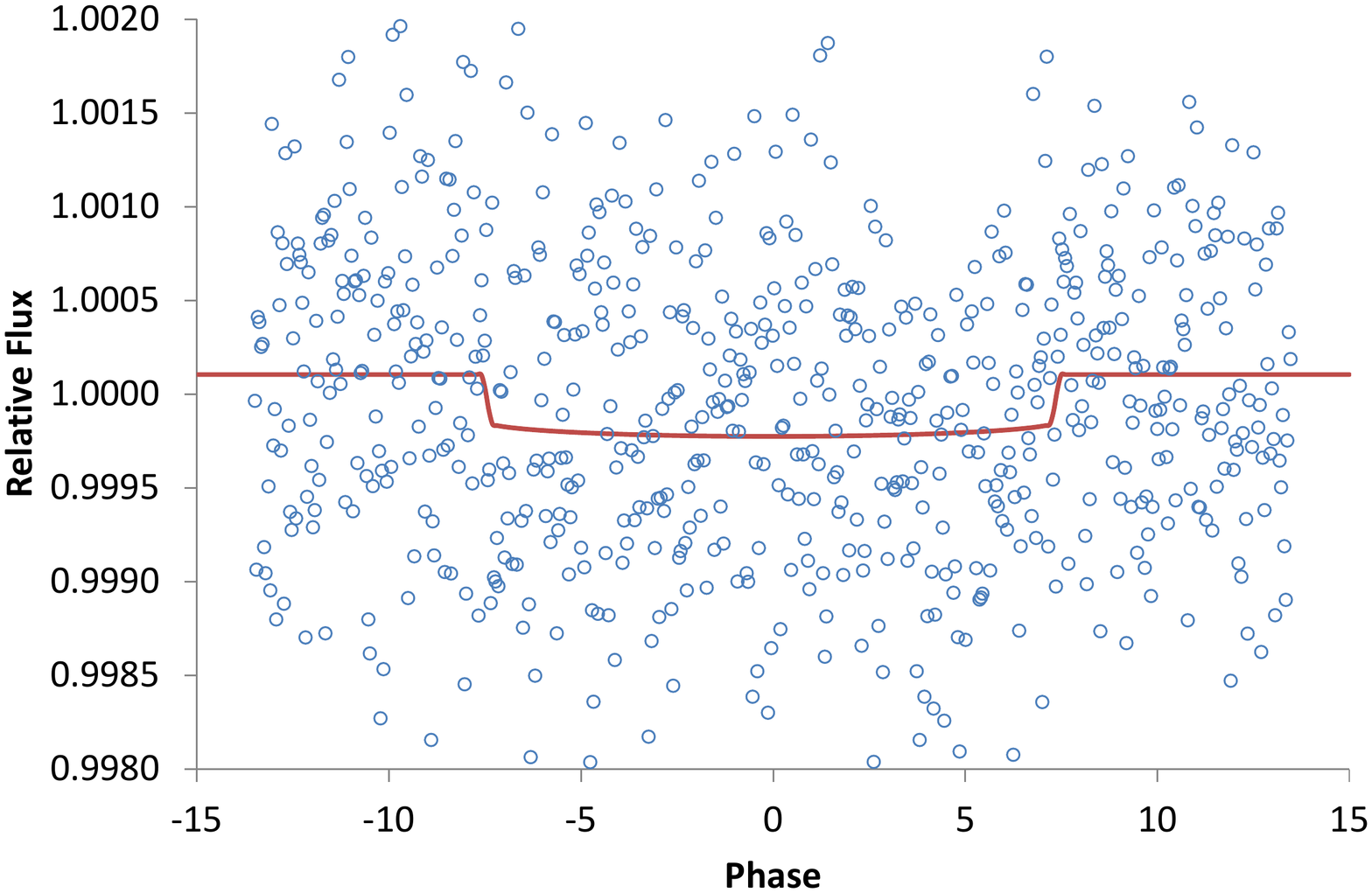} 
\caption{NEA light curve of KOI 388.01, and its {\sc WinKepler} model.}
\label{fig:KOI388}
\end{center}
\end{figure} 

\begin{figure}[H]
\begin{center}
\includegraphics[width=\columnwidth]{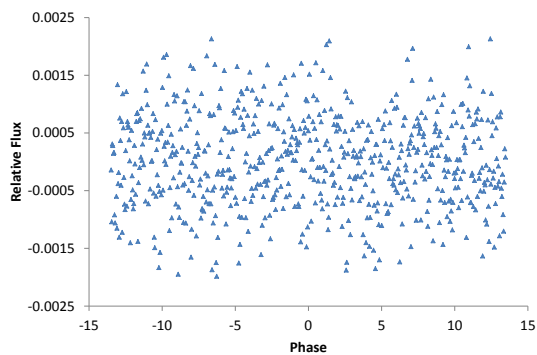} 
\caption{The residuals diagram for KOI 388.01}
\label{fig:KOI388d}
\end{center}
\end{figure} 

\begin{figure}[H]
\begin{center}
\includegraphics[width=\columnwidth]{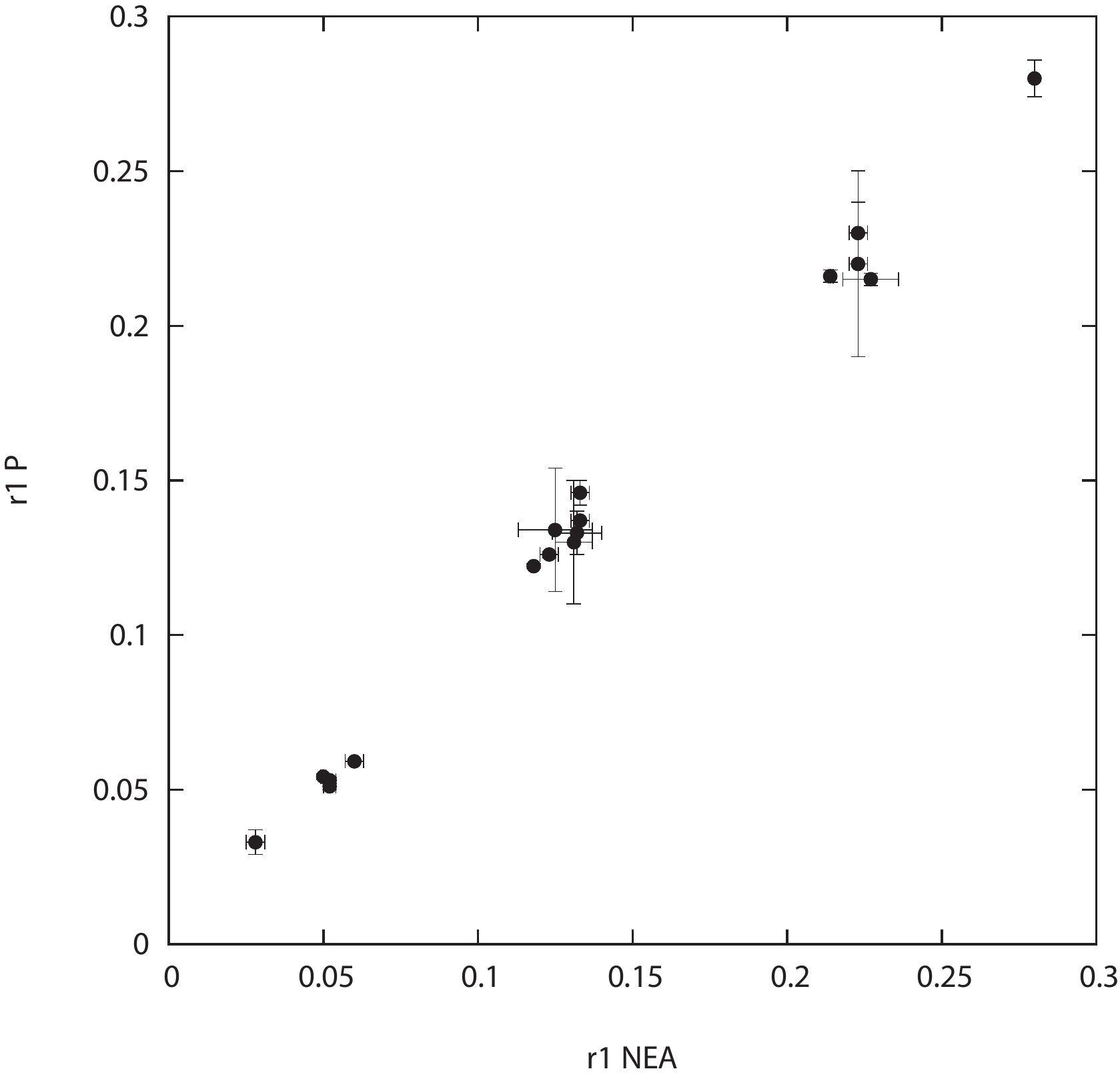} 
\caption{R*/a (K vs P values).  Both K and P error estimates are generally comparable, though the P erros are often somewhat larger. The exception is KOI 13.01, where the NEA listing recognizes possible problems in the modelling. The error measures reflect the deterioration in accurate parameter specification for the less well-defined eclipses, as well as loss of determinacy with more parameters in the set.}
\label{fig:RatioOfR_1}
\end{center}
\end{figure} 

\begin{figure}[H]
\begin{center}
\includegraphics[width=\columnwidth]{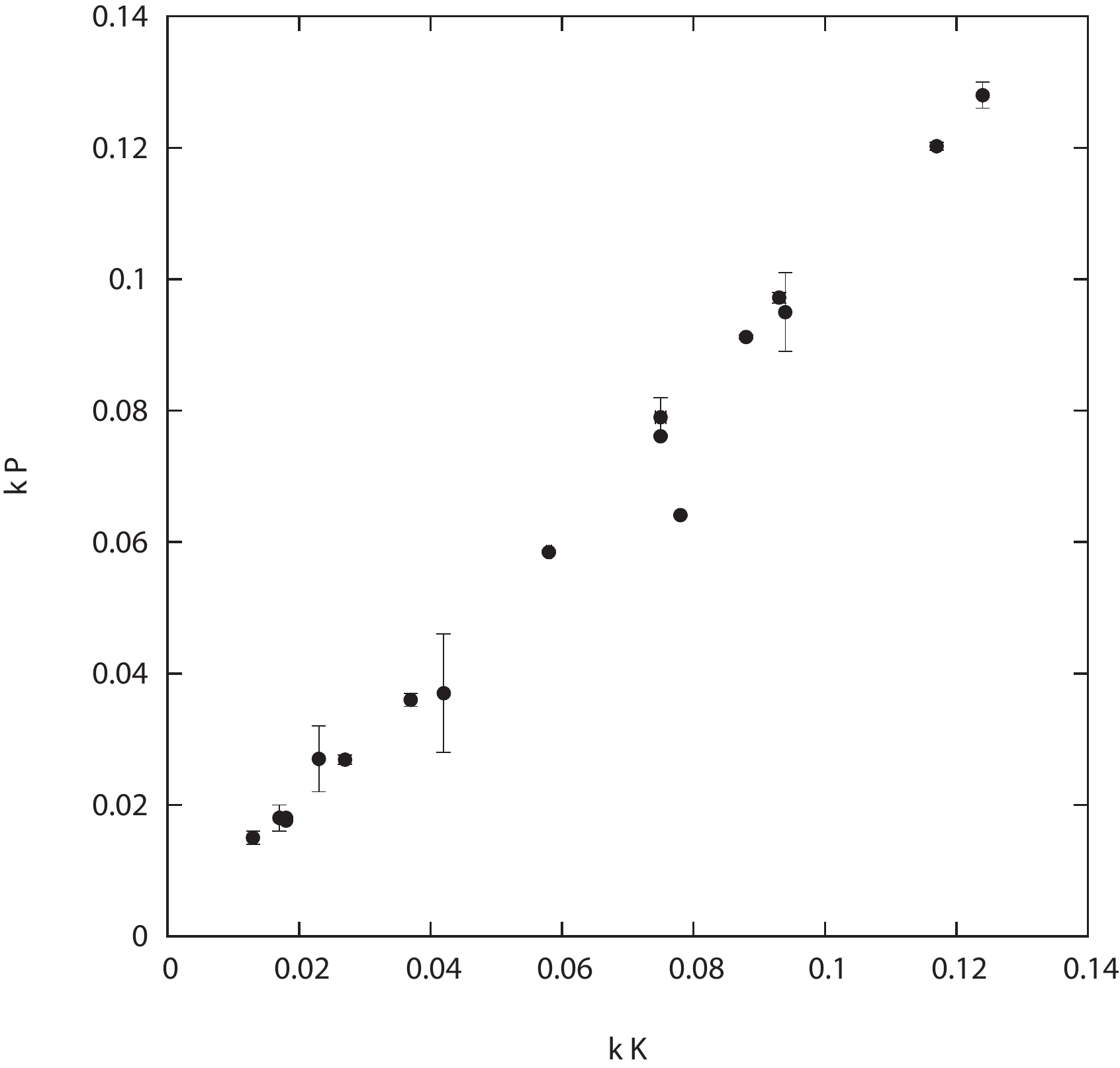} 
\caption{Ratio of Radii (K vs P values). The P error estimates, calculated with full interdependence of the parameters, are often significantly larger than the K ones. The difference in $k$ value for
KOI 13.01, however, significantly exceeds either error estimate and must reflect a modelling difference for this complex example (see Table \ref{tab:no3}).}
\label{fig:RatioOfRadii}
\end{center}
\end{figure} 

\begin{figure}[H]
\flushleft
\includegraphics[width=\columnwidth]{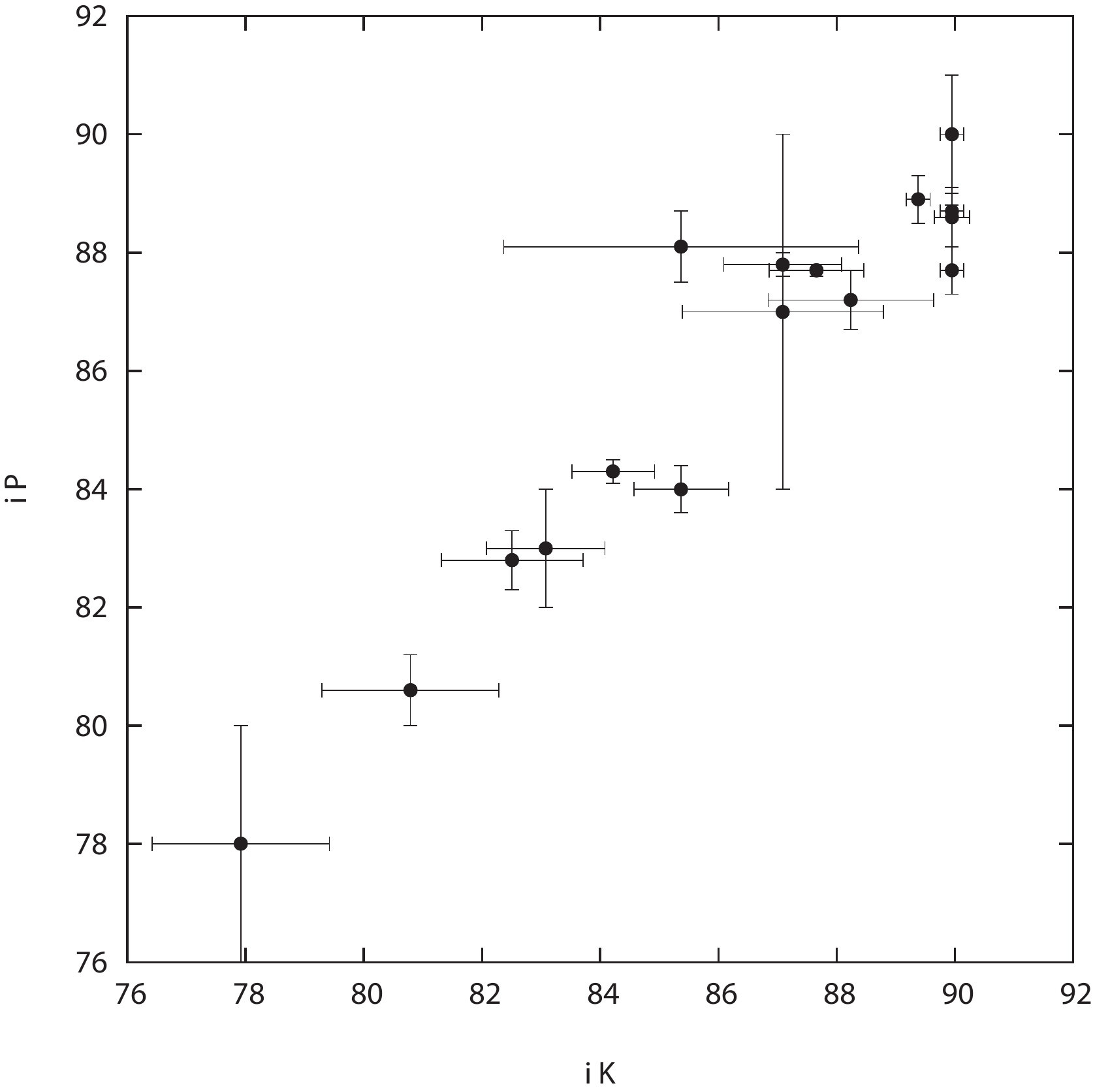}
\caption{Inclinations (K vs P values). The K error estimates are derived from cited errors of the impact parameter $b$, so involve also the errors of the relative radius $r_1$. The results are comparable from both sources in most cases. The difference in the solution for KOI 13.01 is again evident, but the derived K error is sufficiently large as to reduce its significance.}
\label{fig:RatioOfInclination}
\end{figure}  

\end{document}